\documentclass[prl, twocolumn,superscriptaddress,amsmath,amssymb,showpacs,floatfix,preprintnumbers]{revtex4-2}

\usepackage{amsmath}
\usepackage{mathtools}
\usepackage{graphicx}
\usepackage{dcolumn}
\usepackage{xcolor}
\usepackage{physics}
\usepackage{xr}
\usepackage{siunitx}
\usepackage{tabularx}
\usepackage{makecell}
\usepackage{multirow}
\DeclareGraphicsExtensions{.png .jpg .pdf}

\usepackage[normalem]{ulem}

\makeatletter
\newcommand*{\addFileDependency}[1]{% argument=file name and extension
  \typeout{(#1)}
  \@addtofilelist{#1}
  \IfFileExists{#1}{}{\typeout{No file #1.}}
}
\makeatother
 
\newcommand*{\myexternaldocument}[1]{%
    \externaldocument{#1}%
    \addFileDependency{#1.tex}%
    \addFileDependency{#1.aux}%
}

\myexternaldocument{si}

\begin{document}

\title{Moir\'e Collapse and Luttinger Liquids In Twisted Anisotropic Homobilayers}

\author{D. J. P. de Sousa}
\author{Seungjun Lee}
\affiliation{Department of Electrical and Computer Engineering, University of Minnesota, Minneapolis, Minnesota 55455, USA}
\email{sousa020@umn.edu}
\author{Francisco Guinea}
\affiliation{IMDEA Nanoscience, Faraday 9, 28049 Madrid, Spain}
\affiliation{Donostia International Physics Center, Paseo Manuel de Lardizábal 4, 20018 San Sebastián, Spain}
\author{Tony Low}
\affiliation{Department of Electrical and Computer Engineering, University of Minnesota, Minneapolis, Minnesota 55455, USA}
\affiliation{Department of Physics, University of Minnesota, Minneapolis, Minnesota 55455, USA}
\email{tlow@umn.edu}

\begin{abstract}

%Moir\'e quantum matter has emerged as a versatile condensed matter platform, enabling unprecedented control over the electronic structure of van der Waals materials. Despite the vast diversity of currently available anisotropic two-dimensional (2D) materials, it remains unclear whether moir\'e patterns in twisted anisotropic homobilayers exhibit distinctive features beyond the conventional moir\'e system's paradigm. In this work, we introduce twisted homobilayers of anisotropic 2D materials as a new class of moir\'e systems, where the twist angle provides a tuning knob to modulate the system's anisotropy to extreme degrees. By varying the twist angle from $90^{\circ}$ to a critical value $\theta_M$, we demonstrate a continuous transition from \textit{full isotropic} moir\'e patterns towards \textit{unbounded} degrees of anisotropy, characterized by the collapse of the mBZ. By virtue of its geometric origin, these extreme anisotropic configurations near the moir\'e collapse can profoundly impact electronic behavior, providing a means to realize non-Fermi liquids such as Luttinger and sliding Luttinger phases. We systematically explore the consequences of this moir\'e collapse on the band structure of representative twisted anisotropic bilayers, including black phosphorus, SnSe, and GeTe. Furthermore, we develop a continuum model for twisted black phosphorus bilayers that comprehensively captures the moir\'e physics of electrons at large twist angles, encompassing the regime of extreme anisotropies between $90^{\circ}$ and $\theta_M$.

We introduce twisted anisotropic homobilayers as a distinct class of moiré systems, characterized by a distinctive ``magic angle", $\theta_M$, where both the moiré unit cell and Brillouin zone collapse. Unlike conventional studies of moiré materials, which primarily focus on small lattice misalignments, we demonstrate that this moiré collapse occurs at large twist angles in generic twisted anisotropic homobilayers. The collapse angle, $\theta_M$, is likely to give rise quasi-crystal behavior as well as to the formation of strongly correlated states, that arise not from flat bands, but from the presence of ultra-anisotropic electronic states, where non-Fermi liquid phases can be stabilized. In this work, we develop a continuum model for electrons based on extensive \textit{ab initio} calculations for twisted bilayer black phosphorus, enabling a detailed study of the emerging moir\'e collapse features in this archetypal system. We show that the (temperature) stability criterion for the emergence of (sliding) Luttinger liquids is generally met as the twist angle approaches $\theta_M$. Furthermore, we explicitly formulate the collapsed single-particle one-dimensional (1D) continuum Hamiltonian and provide the \textit{fully interacting}, bosonized Hamiltonian applicable at low doping levels. Our analysis reveals a rich landscape of multichannel Luttinger liquids, potentially enhanced by valley degrees of freedom at large twist angles.

\end{abstract}
%\pacs{71.10.Pm, 73.22.-f, 73.63.-b}

\maketitle

\section{Introduction} Moir\'e quantum matter has emerged as a powerful platform for engineering exotic electronic states in two-dimensional (2D) systems and beyond~\cite{carr_twistronics_2017, bistritzer_moire_2011, Kennes2021, lopesdossantos_graphene_2007, cao_unconventional_2018,cao_correlated_2018, PhysRevLett.123.036401, PhysRevLett.129.047601, trambly_de_laissardiere_localization_2010, kapfer_programming_2023, pena_moire_2023, angeli__2021, guinea_continuum_2019, naik_ultraflatbands_2018, huder_electronic_2018, lopes_dos_santos_continuum_2012, li_quasiperiodic_2024, Lee2024, https://doi.org/10.48550/arxiv.2411.16497, https://doi.org/10.48550/arxiv.2409.06806, Slot2023, Yoo2019}. Precise control over the lattice misalignment between atomically thin materials has enabled remarkable tunability of their electronic structures, enabling topological flat-band physics, and the potential to emulate strongly correlated archetypal condensed matter systems~\cite{Kennes2021, PhysRevLett.121.026402, san-jose_non-abelian_2012, Devakul2021, PhysRevLett.129.047601, Wang2022, Yu2023, Sharpe2019, Serlin2020}. Thus far, most efforts have focused on moir\'e systems based on isotropic 2D materials. In contrast, moir\'e quantum matter derived from twisted anisotropic homobilayers — such as those composed of black phosphorus (BP) or other low-symmetry materials~\cite{rudenko_anisotropic_2024, Li2019, lin_multilayer_2016, wu_atomic_2015, rudenko_intrinsic_2016, liu_mobility_2016, ling_renaissance_2015, castellanos-gomez_isolation_2014, xia_rediscovering_2014, tran_layer-controlled_2014, liu_phosphorene_2014, koenig_electric_2014, PhysRevLett.133.146605, Kowalczyk2020, Du2022, PhysRevB.101.184101, Wang2017, Bao2019} — is relatively less explored. Although some progress has been made in recent years~\cite{Srivastava2021, Huang2024, PhysRevB.105.235421, Chen2024, Zhu2023, Xiong2022, Wang2021, 2Wang2022, Wang2022, Yu2023, He2021, Sevik2017, Guo2023, PhysRevB.96.195406, Cao2016, PhysRevB.96.195406}, it remains unclear whether these systems introduce fundamentally new moir\'e phenomena or whether they merely extend the conventional paradigm established for isotropic materials, with additional degrees of freedom stemming from their anisotropic nature. Considering the wide variety of experimentally accessible 2D anisotropic materials, exploring this question could yield important insights into the existence of currently inaccessible exotic quantum states of matter and open pathways for designing next-generation quantum materials with desirable properties for applications in nanoelectronics and optoelectronics~\cite{Kim2016, lin_multilayer_2016, low_topological_2015, Forte2019, PhysRevB.97.155424, van_veen_tuning_2019, low_tunable_2014, chaves_anisotropic_2015, zhang_infrared_2017, arra_exciton_2019, zhang_determination_2018, qiu_environmental_2017, ghosh_anisotropic_2017, villegas_two-dimensional_2016, correas-serrano_black_2016, surrente_excitons_2016, nemilentsau_anisotropic_2016, tran_quasiparticle_2015, wang_highly_2015, low_plasmons_2014}.

%%%%%%%%%%%%%%%%%%%%%%%%%%%%%%%%%%%%%%%%%%%%%%%%%%%%%%%%%%%%%%%%
%
%                   BEGIN: Figures
%
%%%%%%%%%%%%%%%%%%%%%%%%%%%%%%%%%%%%%%%%%%%%%%%%%%%%%%%%%%%%%%%%
\begin{figure*}[t]
\centerline{\includegraphics[width = \linewidth]{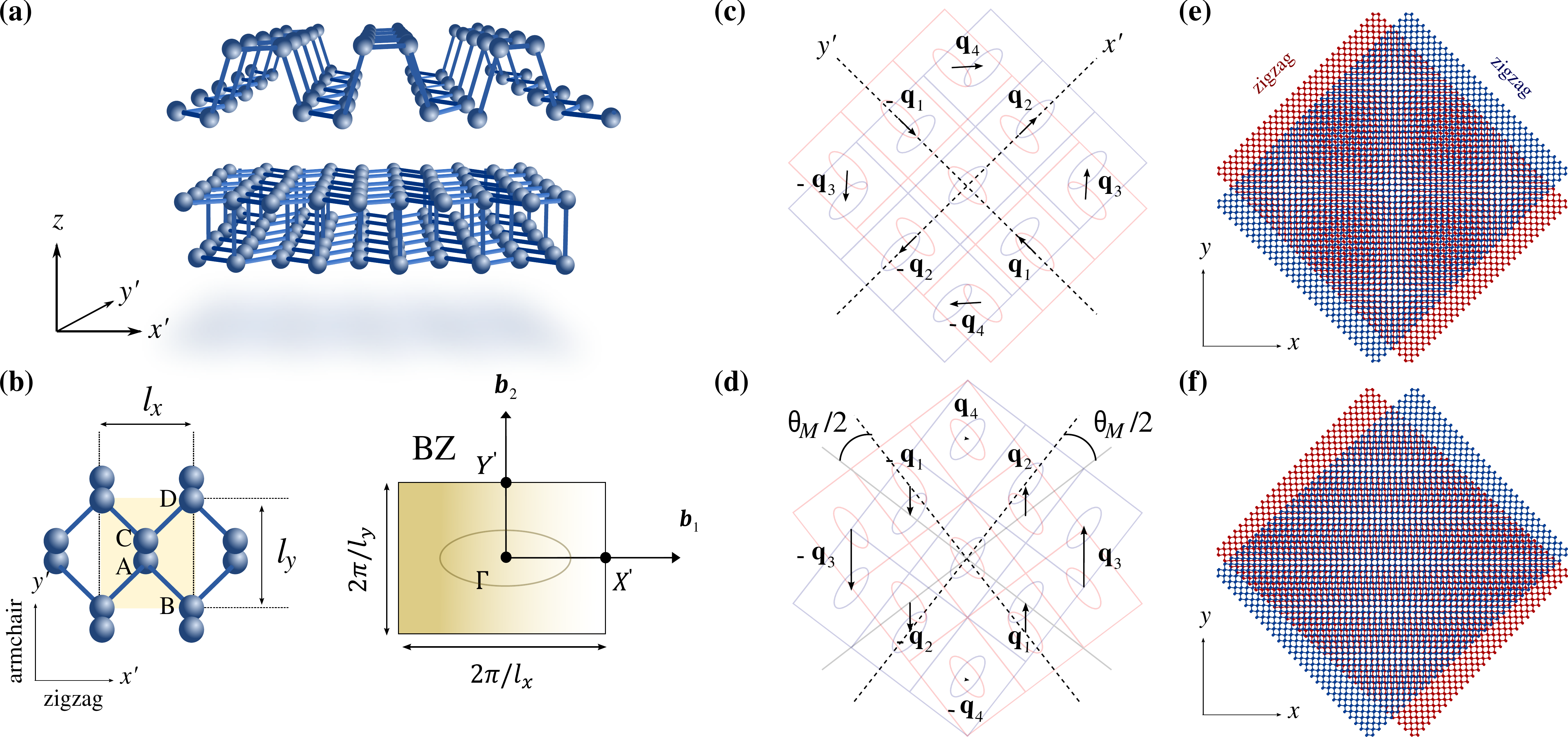}}
\caption{\textbf{Moir\'e collapse in twisted anisotropic homobilayers}. (a) 90$^\circ$ twisted black phosphorus homobilayer. (b) Unit cell (left) and Brillouin zone (right) of monolayer black phosphorus. The anisotropic monolayer unit cell, with dimensions $l_x$ and $l_y$, contains four atomic sites denoted as $A$, $B$, $C$ and $D$. The monolayer Brillouin zone inherits the real space anisotropy. The shaded ellipse represents the Fermi surface centered at the $\Gamma$ point. The distribution of the smallest transfer momenta, $\textbf{q}_{1,2,3,4}$, for the twisted black phosphorus homobilayer is represented in panels (c) and (d). While the transfer momenta span a two-dimensional (2D) space in the $90^{\circ}$ twisted configuration in (c), panel (d) depicts a situation where all transfer momenta align along a particular direction at a finite twist angle, $\theta_M$, configuring a collapse into a one-dimensional (1D) transfer momenta space. Panels (e) and (f) display the moir\'e pattern of a representative twisted anisotropic homobilayer at $90^{\circ}$ and near $\theta_M$. The real space axis for the zigzag and armchair orientations are $x'$ and $y'$. We reserve $x$ and $y$ axis to describe the real space orientation of the moir\'e pattern, as defined above.}
\label{Fig1}
\end{figure*}

\begin{figure*}[t]
\centerline{\includegraphics[width = \linewidth]{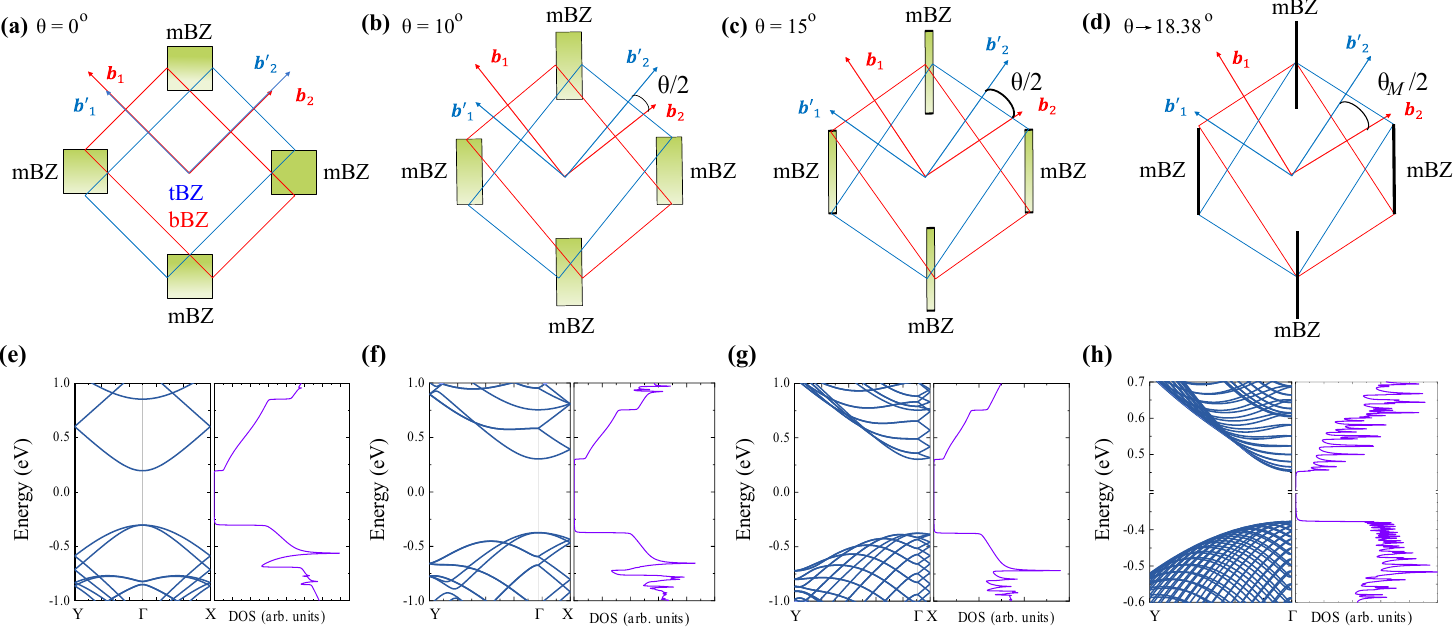}}
\caption{\textbf{The evolution of the electronic structure in twisted black phosphorus homobilayers towards the collapse limit}. (a)-(d) The progression of the moir\'e Brillouin zone in twisted bilayer black phosphorus, starting from the $90^{\circ}$ ($\theta = 0^{\circ}$ in our convention) configuration. The red and blue rectangular Brillouin zones, with reciprocal lattice vectors $\textbf{b}_{1,2}$ and $\textbf{b}'_{1,2}$, respectively, represent the k-space of the bottom and top monolayers of black phosphorus. The moir\'e Brillouin zone, shown in green, is confined within the k-space region between the corners of the top and bottom Brillouin zones, denoted as tBZ and bBZ. As the twist angle $\theta$ increases, the initially isotropic mBZ becomes increasingly compressed until it collapses at a finite angle $\theta_M$. (e)-(h) The evolution of the electronic structure and density of states in twisted bilayer black phosphorus as it approaches the collapse limit. As the twist angle $\theta$ deviates from $0^{\circ}$, the initially isotropic conduction and valence bands begin to develop anisotropy. This leads to dramatic changes in the electronic behavior as $\theta$ approaches $\theta_M$, where the bands are described by a single crystal momentum over the collapsed moir\'e Brillouin zone, and 1D van Hove singularities emerge.}
\label{Fig2}
\end{figure*}

\begin{figure*}[t]
\centerline{\includegraphics[width = \linewidth]{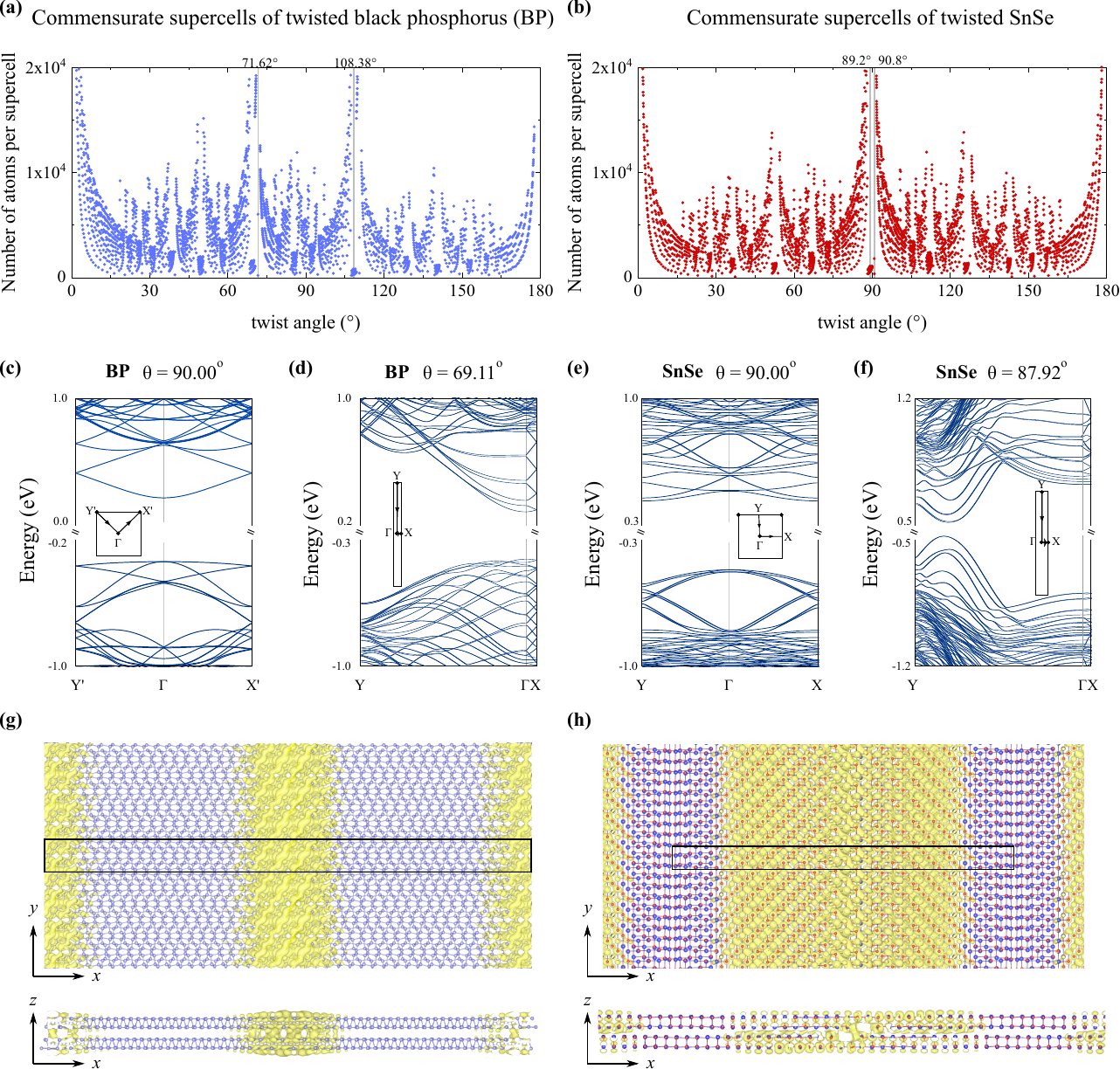}}
\caption{\textbf{The moir\'e collapse through the lens of \textit{ab initio} calculations}. (a) and (b) show the possible choices of commensurate supercells for twisted homobilayer black phosphorus (BP) and SnSe, respectively, highlighting the number of atoms at a given twist angle. The vertical lines at $90^{\circ} \pm \theta_M$ in both plots are the predicted collapse magic angle, closest to the $90^{\circ}$ configuration. The divergent number of atoms per supercell around these lines reflect collapse tendency. (c) and (d) show the \textit{ab initio} electronic structure of homobilayer BP twisted at $90^\circ$ and $69.11^\circ$, respectively. Panels (d) and (f) show the same, but for twisted SnSe homobilayers at $90^\circ$ and $87.92^\circ$. The insets sketch the moir\'e Brillouin zone for each case. The $X'$ and $Y'$ high-symmetry points in panel (c) refer to the zigzag and armchair crystallographic orientations of the individual BP monolayers. Near the collapse, the real space moir\'e unit cell becomes elongated along the $\textbf{a}_{2M}$ direction, with the associated reciprocal space direction considerably shortened ($\Gamma - X$).The $\Gamma$ point charge densities, at the lowest conduction state, for the two nearly collapsed systems are shown in panels (e) and (d), where the quasi-1D feature is evident. The moir\'e unit cells are highlighted.}
\label{Fig3}
\end{figure*}

%%%%%%%%%%%%%%%%%%%%%%%%%%%%%%%%%%%%%%%%%%%%%%%%%%%%%%%%%%%%%%%%
%
%                   END: Figures
%
%%%%%%%%%%%%%%%%%%%%%%%%%%%%%%%%%%%%%%%%%%%%%%%%%%%%%%%%%%%%%%%%

In this work, we introduce a universal and distinct feature of moir\'e systems based on \textit{anisotropic} 2D materials — the \textit{moir\'e collapse}~\cite{PhysRevLett.131.166402}: In these systems, the anisotropy of the moir\'e pattern can be tuned to extreme levels, ultimately leading to a collapsed one-dimensional (1D) crystal at a finite twist angle, $\theta_M$. To demonstrate and explore the consequences of this feature, we develop a continuum model that captures the evolution of the electronic structure of a BP bilayer twisted at large angles, close to $90^\circ$, built on extensive \textit{ab initio} calculations. Although the study focuses on twisted bilayer black phosphorus (TBBP), the collapse angle is found to depend exclusively on the anisotropy of the unit cell in the constituent monolayers, not on the type of atoms and particular basis arrangement. Consequently, we also show that the derived $\theta_M$ successfully predicts the presence of highly anisotropic, quasi-1D moir\'e structures of twisted anisotropic homobilayers made from other materials. 

A striking consequence of the moir\'e collapse is the ability to induce (quasi-) 1D physics with arbitrary precision, determined solely by proximity to $\theta_M$; The presence of a finite $\theta_M$ introduces a structural configuration through which strongly correlated (sliding) Luttinger liquid phases can be stabilized, similar to how the magic angle in twisted bilayer graphene facilitates the formation of correlated phases. Detailed investigation of the evolution of the electronic structure of TBBP toward the (quasi-) 1D limit is presented (see supplementary material for more details at Ref.~\cite{Snote}). We also derive the 1D electronic Hamiltonian at the collapse limit and utilize it to write the 1D fully interacting bosonized Hamiltonian for the collapsed TBBP. The Luttinger parameters and the plasmon dispersion are explicitly calculated, offering a testable prediction for experimental validation.

\section{The Moir\'e Collapse in Twisted Anisotropic Homobilayers}

We begin by describing the evolution of the moir\'e Brillouin zone and the unit cell of twisted anisotropic homobilayers for small misalignment near the $90^{\circ}$ configuration. Without loss of generality, we focus on the description of TBBP as a prime example of anisotropic homobilayer, schematically shown in Fig.~\ref{Fig1}(a). Our convention is that the zigzag direction of the bottom BP layer is aligned with the $x$ axis and the top BP layer is misaligned by $90^{\circ}$. Figure~\ref{Fig1}(b) highlights the anisotropic unit cell, sublattices (A, B, C and D) and the Brillouin zone of the bottom BP monolayer. In this work, we define the twist angle $\theta$ as measured from this initial $90^{\circ}$ twisted configuration. 

The reciprocal lattice vectors of the moir\'e superlattice are defined from the smallest transfer momenta, $\textbf{q}_j$ ($j = 1,2,3,4$), consistent with the symmetries of the moir\'e pattern~\cite{Snote}. The smallest vectors are defined as $\textbf{q}_1 = \textbf{b}_1 - \textbf{b}'_1$, $\textbf{q}_2 = \textbf{b}_2 - \textbf{b}'_2$, $\textbf{q}_3 = \textbf{q}_2 + \textbf{q}_1$ and $\textbf{q}_4 = \textbf{q}_2 - \textbf{q}_1$, where $\textbf{b}'_1 = (2\pi/l_x)\hat{\textbf{x}}$, $\textbf{b}'_2 = (2\pi/l_y)\hat{\textbf{y}}$ are the reciprocal lattice vectors of the bottom BP layer and $\textbf{b}_1 = (2\pi/l_y)\hat{\textbf{x}}$, $\textbf{b}_2 = (2\pi/l_x)\hat{\textbf{y}}$ correspond to the reciprocal lattice vector of the top layer, in the 90$^{\circ}$ configuration. The transfer momenta are schematically shown in Fig.~\ref{Fig1}(c), where we sketch the momentum space of the top (bottom) BP layer in red (blue). The structure of the moir\'e reciprocal space is dictated by all possible reciprocal lattice vectors, $\textbf{G}$, defined in the lattice formed by the two linearly independent transfer momenta $\textbf{q}_3$ and $\textbf{q}_4$, i.e., $\textbf{G} = \{ n\textbf{b}_{1M} + m\textbf{b}_{2M} | n,m \in \mathbb{Z} \}$, with $\textbf{b}_{1M} = \textbf{q}_3/2$ and $\textbf{b}_{2M} = \textbf{q}_4/2$ being the reciprocal lattice vector of the moir\'e superlattice consistent with a rectangular unit cell symmetry. 

As the angle between the two BP layers deviates from $\theta = 0^\circ$ (away from the $90^{\circ}$ configuration), all transfer momenta gradually align along a particular direction in momentum space. We find that this process unavoidably results in full alignment of all transfer momenta at a finite critical twist angle, $\theta_{M}$, as illustrated in Fig.~\ref{Fig1}(d). In the limit $\theta \rightarrow \theta_M$, the moir\'e reciprocal lattice basis vectors approach $\textbf{b}_{2M} \rightarrow \textbf{0}$ and $\textbf{b}_{1M} \rightarrow \textbf{b}_{1M}(\theta_M)$, such that the moir\'e reciprocal space becomes 1D, with reciprocal lattice vectors taking values in the 1D lattice $\textbf{G} = \{ n\textbf{b}_{1M}(\theta_M)\}$, with integer $n$. This represents a complete collapse of the 2D moir\'e superlattice into a 1D crystal.

In real space, the moir\'e lattice vectors assume the form
\begin{subequations}
\begin{eqnarray}
\textbf{a}_{1M} = \displaystyle \frac{\bar{l}(\theta)}{\tan(\theta_M/2) + \tan(\theta/2)} (\hat{\textbf{x}} - \hat{\textbf{y}}), 
    \label{eq2a}
\end{eqnarray}
\begin{eqnarray}
\textbf{a}_{2M} = \displaystyle \frac{\bar{l}(\theta)}{\tan(\theta_M/2) - \tan(\theta/2)} (\hat{\textbf{x}} + \hat{\textbf{y}}), 
    \label{eq2b}
\end{eqnarray}
\end{subequations}
sharing a common multiplicative factor $\bar{l}(\theta) = [(1/l_x + 1/l_y)\cos(\theta/2)]^{-1}$. As the twist angle progressively approaches the collapse limit, $\theta \rightarrow \theta_M$, the primitive moir\'e lattice vectors approach $\textbf{a}_{1M} \rightarrow [2\sin(\theta_M/2)(1/l_x + 1/l_y)]^{-1}$ and $\textbf{a}_{2M} \rightarrow \infty$. This result indicates that while a periodicity can be established for the moir\'e pattern along the $\textbf{a}_{1M}$ direction in real space, such periodicity cannot be defined along the $\textbf{a}_{2M}$ direction. The 1D collapsed moir\'e system exhibits a complete lack of commensurability between the two layers in the $\textbf{a}_{2M}$ direction. Consequently, the 2D arrangement of atoms at the moir\'e collapse effectively constitutes a 1D crystal with an infinitely extended atomic basis along the $\textbf{a}_{2M}$ direction. The real space moir\'e patterns, at $\theta = 0^{\circ}$ and $\theta \approx \theta_M$, of a representative anisotropic homobilayer are shown in Figs.~\ref{Fig1}(e) and (f).

We emphasize that the moir\'e collapse discussed above is a distinctive feature of \textit{large} angle moir\'e patterns in twisted anisotropic homobilayers. While the anisotropy of the moir\'e unit cell can be finely tuned to extreme levels near the $90^{\circ}$ configuration, the anisotropy of the moir\'e pattern at small twist angles is \textit{fixed}; To quantify this behavior, we define the anisotropy of the moir\'e unit cell by $\zeta(\theta) = |\textbf{a}_{1M}/\textbf{a}_{2M}|$, and we find $\zeta(\theta) = [\tan(\theta_M/2) + \tan(\theta/2)]/[\tan(\theta_M/2) - \tan(\theta/2)]$, when $\theta$ measures the misalignment from the $90^{\circ}$ configuration and $\zeta(\theta) = l_x/l_y$ when $\theta$ measures the misalignment from the AB-stacked configuration~\cite{Snote}. The first collapse angle $\theta_M$, closest to the $90^{\circ}$ configuration, for general anisotropic twisted homobilayers is predicted to depend only on $l_x$ and $l_y$ via 
\begin{eqnarray}
\theta_M = \displaystyle 2 \tan^{-1}\left(\frac{l_y - l_x}{l_y + l_x}\right).
    \label{eq1}
\end{eqnarray}
For the TBBP, with $l_x = 0.3294$ nm and $l_y = 0.4565$ nm, it is predicted that the collapse of the mBZ will take place as $\theta_{M} \rightarrow 18.38^{\circ}$ off 90$^{\circ}$ ($\theta_{M} \rightarrow 71.62^{\circ}$ off 0$^{\circ}$).

Research on moi\'e quantum matter in anisotropic systems has largely focused on emergent phenomena at small twist angles (near $0^\circ$ misalignment)~\cite{Wang2021, 2Wang2022, Wang2022, Yu2023}. As a result, the phenomenon of moir\'e collapse and its broader implications have been overlooked. The moir\'e collapse described above is expected to influence all types of excitations in twisted anisotropic homobilayers, including phonons, excitons, plasmons, and electrons. Furthermore, the mechanical properties of these systems, such as elastic constants, are anticipated to display critical behavior near the moir\'e collapse, offering alternative avenues to probe this phenomenon. In the following section, we will focus on examining the effect of the moir\'e collapse specifically on the behavior of electrons.

\section{Electrons and Luttinger Liquids at the Moir\'e Collapse}

An intriguing outcome of the moir\'e collapse discussed in the previous section is the emergence of (quasi-) 1D electronic behavior. Ideal 1D interacting electron systems are especially fascinating due to the emergence of Luttinger liquid phases~\cite{Tomonaga1950, Luttinger1963, RevModPhys.84.1253, Haldane1981, PhysRevLett.45.1358, 2Haldane1981, book, PhysRevLett.68.1220, Jompol2009, PhysRevB.82.245104}. Close to the moir\'e collapse, quasi-1D physics could also lead to the emergence of sliding Luttinger liquids (SLL), a non-Fermi liquid phase resulting from the lateral coupling of 1D Luttinger wires~\cite{PhysRevLett.86.676, PhysRevB.64.045120}. Recent experiments have proposed the emergence of SLL behavior in twisted 1T'-WTe$_{2}$ bilayers~\cite{Yu2023, Wang2022}. These findings were restricted to moir\'e states at small twist angles, where the anisotropy of the moir\'e unit cell cannot be modified. Here, we demonstrate that at large twist angles - where the anisotropy is highly tunable - the conditions become more favorable for stabilizing strongly correlated quasi-1D phases, with ``ideal” Luttinger liquids forming in close proximity to the moir\'e collapse.

In this work, we develop a continuum model that captures the evolution of the electronic structure of TBBP from the $90^{\circ}$ configuration to the collapse limit. The model is constructed over the low-energy electron states within each BP monolayer. The two relevant basis orbitals correspond to symmetric superpositions of the $p_z$ orbitals at the A and D sites (referred to as the $\phi_{AD}$ orbitals) and at the B and C sites (referred to as the $\phi_{BC}$ orbitals)~\cite{PhysRevB.96.155427, PhysRevB.92.075437}. The Hamiltonian for TBBP follows the structure of the Bistritzer and MacDonald model~\cite{bistritzer_moire_2011, PhysRevB.81.245412} of twisted bilayer graphene, but with all parameters extracted from first-principles calculations. The complete derivation can be found in the accompanying supplementary materials~\cite{Snote}. 

Denoting the plane-wave annihilation (creation) operator as $c_{\textbf{k}\textbf{G}\alpha\l}$ ($c^{\dagger}_{\textbf{k}\textbf{G}\alpha\l}$), where $\textbf{k}$ is the crystal momentum within the mBZ, $\alpha = AD, BC$ is the orbital index and $l = t, b$ is the layer index, the second quantized single particle Hamiltonian in the basis $c^{\dagger}_{\textbf{k}\textbf{G}} = [c_{\textbf{k}\textbf{G},AD,t} \ \ c_{\textbf{k}\textbf{G},BC, t} \ \ c_{\textbf{k}\textbf{G},AD, b}  \ \ c_{\textbf{k}\textbf{G},BC, b}]^{\dagger}$ reads
\begin{eqnarray}
\hat{H} = \displaystyle \sum_{\textbf{k}\in \textrm{MBZ}}\sum_{\textbf{G}\textbf{G}'} H_{\textbf{G}\textbf{G}'}(\textbf{k}) c^{\dagger}_{\textbf{k}\textbf{G}}c_{\textbf{k}\textbf{G}'}.
    \label{secV:eq27}
\end{eqnarray}
with plane-wave matrices
\begin{eqnarray}
H_{\textbf{G}\textbf{G}'}(\textbf{k}) = [H_0(\textbf{k} + \textbf{G}) + T_0] \delta_{\textbf{G}\textbf{G}'} + \nonumber \\ \sum_{j = 1,2}T_j(\delta_{\textbf{G} - \textbf{G}', \textbf{q}_j} + \delta_{\textbf{G} - \textbf{G}', -\textbf{q}_j}).
    \label{secV:eq29}
\end{eqnarray}
Here, $H_0(\textbf{k})$ corresponds to the bare uncoupled bilayer Hamiltonian, whereas $T_{j = 0,1,2}$ describes the interlayer moir\'e coupling. The uncoupled bilayer Hamiltonian reads
\begin{eqnarray}
     H_{0}(\textbf{k}) = \left[
     \begin{tabular}{cc}
      $h_{0}[R(\pi/2+\theta/2)\textbf{k}]$  &  $0$ \\
       $0$ &  $h_{0}[R(-\theta/2)\textbf{k}]$
     \end{tabular}\right], \label{secV:eq30}
\end{eqnarray}
where $R(\theta)$ is the rotation operator. The bare monolayer black phosphorus continuum Hamiltonian is~\cite{PhysRevB.96.155427, PhysRevB.94.235415, PhysRevB.92.075437, PhysRevB.89.201408, PhysRevB.92.085419} 
\begin{eqnarray}
     h_{0}(\textbf{k}) = \left[
     \begin{tabular}{cc}
      $u_0 + \eta_x k_x^2 + \eta_y k_y^2$  &  $\delta + \gamma_x k_x^2 + \gamma_y k_y^2 + i\chi k_y$ \\
       $\delta + \gamma_x k_x^2 + \gamma_y k_y^2 - i\chi k_y$ &  $u_0 + \eta_x k_x^2 + \eta_y k_y^2$
     \end{tabular}\right]. \nonumber \\ \label{secV:eq32}
\end{eqnarray}
We found that the minimal set of interlayer moir\'e coupling matrices reflecting the symmetries of the twisted structure are
\begin{eqnarray}
      T_{0} = \left[
     \begin{tabular}{cc}
       $0$ &  $w_0(\sigma_0 + \sigma_x)$ \\
       $w_0(\sigma_0 + \sigma_x)$ &  $0$
     \end{tabular}\right], \nonumber \\  T_{1,2} = \left[
     \begin{tabular}{cc}
       $0$ &  $w_2(\sigma_0 - \sigma_x)$ \\
       $w_2(\sigma_0 - \sigma_x)$ &  $0$
     \end{tabular}\right], \label{secV:eq31}
\end{eqnarray}
where the $\sigma$ matrices operate in the orbital subspace of each monolayer. All model parameters were determined by analyzing the electronic structure of the lowest energy states at twist angles near the $90^{\circ}$ configuration, based on \textit{ab initio} calculations~\cite{Snote}. All continuum model parameters were then obtained by means of a non-linear least square fitting procedure. We refer the interested reader to the supplementary materials for more details~\cite{Snote}. The $\theta$-independent parameters were found to be: $\eta_x = 0.44$ eV $\cdot$ \AA$^2$,  $\eta_y = -1.45$ eV $\cdot$ \AA$^2$, $\gamma_x = 2.18$ eV $\cdot$ \AA$^2$, $\gamma_y = -3.20$ eV $\cdot$ \AA$^2$ and $\chi = 4.82$ eV $\cdot$ \AA. The $\theta$-dependent parameters are: $u_0(\theta) = 0.096 - (0.02/85) * \theta $ eV, $\delta(\theta) = 0.43 + (0.02/85) * \theta $ eV, $w_0(\theta) = 0.055 - (0.017/85) * \theta $ eV and $w_1(\theta) = 0.040 + (0.008/85) * \theta $ eV. 

Next, we investigate the evolution of the reciprocal space and electronic structure of TBBP towards the collapse limit. Additional details on the \textit{ab initio} calculations will be provided in the later sections.

\subsection{The Evolution of the Electronic Structure}

We present the evolution of the mBZ and the electronic structure of the TBBP in Fig.~\ref{Fig2}. The blue and red rectangles represent the Brillouin zones of the top and bottom BP monolayers (tBZ and bBZ) in Figs.~\ref{Fig2}(a)-(d). The geometrical region defined by their corners establishes the shape and dimensions of the mBZ, shown in green in Figs.~\ref{Fig2}(a)-(d). As $\theta$ increases, the initially isotropic mBZ is severely distorted until it collapses to a 1D line at $\theta_M$. The area of the mBZ evolves as $\Omega_{\textrm{BZ}} = \Omega^0_{\textrm{BZ}} [\tan^{2}(\theta_M/2) - \tan^{2}(\theta/2)]$, with $ \Omega^0_{\textrm{BZ}}= 2[\pi \cos(\theta/2) (1/l_x + 1/l_y)]^2$.

The corresponding evolution of the electronic structure is shown in Figs.~\ref{Fig2}(e)-(h). At $\theta = 0^{\circ}$, the in-plane anisotropy of each monolayer is compensated by the adjacent layer, giving rise to isotropic energy bands. Moreover, we found that the presence of a $C_{4z}M_{xy}$ symmetry~\cite{Snote} in this configuration, where $C_{4z}$ is a rotation of $90^{\circ}$ around the $z$ axis and $M_{xy}$ is a layer-interchange operation, prevents the appearance of interlayer hybridization gaps for the valence states at the $\Gamma$ point, as shown in Fig.~\ref{Fig2}(e). As a result, the electronic structure exhibits strongly hybridized conduction states (with a hybridization gap of $2w_0 \approx 110$ meV) alongside unhybridized valence states (with a hybridization gap of $w_2 - w_2 = 0$ meV) near the $\Gamma$ point. These characteristics are consistent with recent experimental observations of a resonant tunneling behavior of holes in twisted black phosphorus homojunctions~\cite{Srivastava2021}, as well as the finite dipole moment of excitons in these structures~\cite{Huang2024}. We note that the opposite occurs at the zone boundaries $X$ and $Y$, where the valence states hybridize, producing van Hove peaks in the density of states, while the conduction states do not hybridize. 

As $\theta$ departs from $0^{\circ}$, the balance between the anisotropy of the two layers is disrupted, resulting in anisotropic electron states. This is shown in Figs.~\ref{Fig2}(f)-(h). The reduction in the size of the mBZ along $\Gamma-X$ effectively increases the number of low-energy states within the considered energy range, reflecting the larger number of orbitals in the elongated real-space moir\'e unit cell. As $\theta$ approaches the collapse limit $\theta_M$, the number of orbital degrees of freedom rises significantly, and the characteristic features of nearly collapsed electron states begin to form: The valence bands display a spectrum akin to that of electrons in a ring geometry subjected to a magnetic flux. This behavior arises from the elongated unit cell near collapse and the periodic boundary condition of the Bloch wave function along the transverse direction. Specifically, the system can be viewed as a nanotube with its axis aligned along the elongated moiré unit cell direction, pierced by a magnetic flux corresponding to the phase acquired by the Bloch wave function for each moir\'e reciprocal lattice vector $G$ along the $\Gamma - X$ direction~\cite{Snote}. The low-energy conduction states acquire quasi-1D multichannel features due to strong interlayer hybridization, with ideal parabolic bands near the $\Gamma$ point. 

As $\theta$ approaches the collapse limit, 1D van Hove singularities begin to emerge, and the system starts to exhibit 1D behavior, as shown in Fig.~\ref{Fig2}(h). We emphasize that the moiré collapse is a limit, meaning our calculations assume that $\theta$ can approach, but never exactly reach, $\theta_M$. Therefore, the nearly collapsed system is characterized by a near-continuum of bands, reflecting the large number of orbital degrees of freedom in the highly anisotropic real-space unit cell. In this limit, the 1D Hamiltonian can be derived as 
\begin{eqnarray}
    H_{GG'}(k) = [H_0(k, G) + T_0]\delta_{GG'},
    \label{secVI:eq38}
\end{eqnarray}
where $H_0 (k, G) = \textrm{diag}[ h^T_G(k)\ \ h^B_G(k)]$. The crystal momentum is now a single number, $k$, taking values in the 1D Brillouin zone of length $\Omega_{\textrm{1D-mBZ}} = |\textbf{b}_{1M}(\theta_M)|$. The relevant matrices are
\begin{eqnarray}
     h_G^T(k) = \left[
     \begin{tabular}{cc}
      $u^T_G + \mu^T_G k + \eta_M k^2$  &  $\delta^T_G + \zeta^T_Gk + \gamma_M k^2 $ \\
        $[\delta^T_G + \zeta^T_Gk + \gamma_M k^2]^{*} $ & $u^T_G + \mu^T_G k + \eta_M k^2$ 
     \end{tabular}\right], \nonumber \\ \label{secV:eq34}
\end{eqnarray}
\begin{eqnarray}
      h_G^B(k) = \left[
    \begin{tabular}{cc}
      $u^B_G + \mu^B_G k + \bar{\eta}_M k^2$  &  $\delta^B_G + \zeta^B_Gk + \bar{\gamma}_M k^2 $ \\
        $[\delta^B_G + \zeta^B_Gk + \bar{\gamma}_M k^2]^{*} $ & $u^B_G + \mu^B_G k + \bar{\eta}_M k^2$ 
     \end{tabular}\right], \nonumber \\ \label{secV:eq35}
\end{eqnarray}
for states around the $\Gamma$ point. Here, we have introduced the functions on the reciprocal lattice vectors, $G = M |\textbf{b}_{M2}|$ ($M = 0, \pm 1, \pm2, ...$), as: $u^T_G = u_0 + \bar{\eta}_M G^2$, $\delta^T_G = \delta + \bar{\gamma}_M G^2 + i\chi_M^x G$, $\mu^T_G = -\Delta \eta_M G$ and $\zeta^T_G = -\Delta \gamma_M G - i\chi^y_M$, for the top layer Hamiltonian, and $u^B_G = u_0 + \eta_M G^2$, $\delta^B_G = \delta + \gamma_M G^2 + i\chi_M^y G$, $\mu^B_G = \Delta \eta_M G$ and $\zeta^B_G = \Delta \gamma_M G + i\chi^x_M$ for the bottom layer Hamiltonian. The parameters are $\eta_M = -0.7983$ eV$\cdot$ \AA$^2$, $\gamma_M = -1.3564$ eV$\cdot$ \AA$^2$, $\chi^y_M = 3.9088$ eV$\cdot$ \AA, $\Delta \gamma_M = -5.1104$ eV$\cdot$ \AA$^2$,  $\bar{\eta}_M = -0.2027$ eV$\cdot$ \AA$^2$, $\bar{\gamma}_M = 0.3414$ eV$\cdot$ \AA$^2$, $\chi^x_M = 2.8203$ eV$\cdot$ \AA, and $\Delta \eta_M = -1.79274$ eV$\cdot$ \AA$^2$. The band structure of Fig.~\ref{Fig2}(h) is obtained from Eq.~(\ref{secVI:eq38}) assuming $|\textbf{b}_{2M}| \approx 2\pi/L$, with $L = 50$ nm. 

Given these salient features, we next analyze the emergence of Luttinger liquid behavior in twisted anisotropic homobilayers.

\subsection{Emergent Luttinger Liquids near the Moir\'e Collapse}

The above results indicate that ideal 1D electron liquids may emerge at the moir\'e collapse. Close to $\theta_M$, the formation of weakly coupled parallel 1D channels (wires) may facilitate the development of sliding Luttinger liquid phases~\cite{PhysRevB.64.045120, Yu2023, PhysRevB.110.L201106, PhysRevB.42.6623, PhysRevLett.86.676,shavit_quantum_2024}. The transition between these phases depends on the strength of the system's anisotropy, that can be approximately captured by defining the effective intra- and inter-wire hoppings constants, $t_{1M}$ and $t_{2M}$, respectively~\cite{PhysRevB.110.L201106}. Assuming that only the lowest conduction band is filled in Fig.~\ref{Fig2}(h), we estimate the hopping constants as $t_{1M} = \hbar^2/2m_{1M}|\textbf{a}_{1M}|^2$ and $t_{2M} = \hbar^2/2m_{2M}|\textbf{a}_{2M}|^2$, where $m_{1(2)M}$ is the effective mass of the lowest conduction band along the corresponding direction. The effective anisotropy is estimated to evolve as,
\begin{eqnarray}
\displaystyle \frac{t_{2M}}{t_{1M}} = \frac{m_{1M}}{m_{2M}} \left[\frac{\tan(\theta_M/2) - \tan(\theta/2)}{\tan(\theta_M/2) + \tan(\theta/2)}\right]^2,
    \label{eqanisotropy}
\end{eqnarray}
indicating that the coupling between 1D channels can be made arbitrarily small as one approaches the moir\'e collapse, i.e., $\lim_{\theta \rightarrow \theta_M} (t_{2M}/t_{1M}) \rightarrow 0$. Therefore, we conclude that it is generally possible to surpass the temperature lower bound, $T_c$, for the emergence of sliding Luttinger liquid behavior by approaching the collapse angle as closely as needed. Following Ref.~\cite{PhysRevB.110.L201106}, the critical temperature is estimated as
\begin{eqnarray}
    k_B T_c \approx t_{2M}\left(\frac{t_{2M}}{t_{1M}}\right)^{\frac{\Delta}{1-\Delta}} \propto \left(\frac{\tan(\theta_M/2) - \tan(\theta/2)}{\tan(\theta_M/2) + \tan(\theta/2)}\right)^{\frac{2}{1-\Delta}}, \nonumber \\
\end{eqnarray}
assuming the anomalous exponent $\Delta$ to lie in the interval $0 \leq \Delta < 1$, where the phase is stabilized down to $T = 0$ for $\Delta > 1$. The upper bound for the temperature is~\cite{book} $k_BT_0 \approx t_{1M}$, such that the 1D Luttinger behavior requires $T < T_0$ as one approaches the collapse. Hence, at sufficiently low temperatures (below $T_0$) strongly correlated Luttinger liquids are expected to emerge at the moir\'e collapse.

To address the properties of these non-Fermi liquids, we proceed by writing the electronic Hamiltonian for interacting electrons at the lowest conduction band of TBBP at $\theta_M$. The key parameter is the Fermi velocity, $v_F$, in the linearized limit. For the lowest conduction state shown in Fig.~\ref{Fig2}(h) at $\epsilon_F \approx 2.8$ meV above the bottom of the band, we find $v_F \approx 5.6\times 10^5$ m/s~\cite{Snote}. Assuming only forward scattering processes, the fully interacting bosonized Hamiltonian is
\begin{eqnarray}
\hat{H} = \displaystyle \frac{1}{2} \hbar u \int dx \left[ K (\partial_x \theta(x))^2 + \frac{1}{K} (\partial_x \phi(x))^2\right],
    \label{bosonized_Hamiltonian}
\end{eqnarray}
for $q \ll k_F\approx 0.068$ nm$^{-1}$. Here, $\phi(x)$ and $\theta(x)$ are the bosonic phase fields and $[\phi(x), -\partial_{x'} \theta(x')] = i\delta(x - x')$ is the canonical commutation relation. By defining $\tilde{g}_4(q) =\lim_{q\rightarrow 0} V(q)/2\pi\hbar v_F$ and $\tilde{g}_2(q) = \lim_{q\rightarrow 0} [V(q) - V(2k_F)]/2\pi\hbar v_F$, we define the Luttinger parameter $K = \sqrt{u_N/u_J}$ and the velocity of the plasmon modes $u = \sqrt{u_N u_J}$ in terms of $u_N = v_F(1 + \tilde{g}_4 - \tilde{g}_2)$ and $u_J = v_F(1 + \tilde{g}_4 + \tilde{g}_2)$. Within our approximations we estimate $K \approx 0.3$ for the collapsed BP. The description presented here should be viewed as the starting point for a sliding Luttinger liquid description of coupled 1D Luttinger ``wires", described by the Hamiltonian~(\ref{bosonized_Hamiltonian}) and laterally coupled through twist angle-dependent ``interwire" hopping $t_{2M}(\theta)$~\cite{PhysRevLett.86.676, PhysRevB.64.045120}. 

We highlight that prominent power-law behavior of the conductance should also manifest near moir\'e collapse, where the transverse conductivity, along $\textbf{a}_{2M}$, scales as~\cite{PhysRevB.110.L201106}: $\sigma_{2M} \propto V^{2\Delta - 1} (T^{2\Delta - 1})$ for $V > k_B T/e$ ( $V < k_B T/e$) for applied bias voltage $V$, where the anomalous exponent is explicitly given as $\Delta = (K + K^{-1} - 2)/4$. We estimate $\Delta \approx 0.41$ from the parameters of the continuum model. We emphasize that 1D Luttinger plasmons and anomalous transport exponents have been experimentally realized systems displaying quasi-1D electronic behavior~\cite{Shi2015, Wang2020, Yu2023}. These experimental advances could enable direct measurements of the moir\'e collapse and emerging strongly correlated electronic behavior using currently available techniques.

At higher electron doping, a few electronic subbands can be populated simultaneously [See the Fig.~\ref{Fig2}(h)], giving rise to multichannel 1D Luttinger liquids. In this regime, the 1D ``Fermi surface" is composed of a few pairs of Fermi points, for each filled subband, where electron states are characterized by distinct velocities. For sufficiently small $q$, the system displays multiple 1D plasmon branches with different dispersions relations. However, at higher $q$ values, a comprehensive theory that incorporates interbranch processes needs to be developed.

The multichannel one dimensional phase outlined here is characterized by the coexistence of at least two incommensurate length scales, coming from the collapse of the two dimensional Brillouin Zone. If electron-electron interactions are neglected, the resulting hamiltonian shares many features with Harper´s equation\cite{harper_single_1955,sokoloff_band_1981}, and other quasi-one-dimensional hamiltonians which include incommensurate length scales, like the Hofstadter model\cite{hofstadter_energy_1976} (see also\cite{sinner_strain-induced_2023}).

While we have primarily focused on TBBP thus far, the key features discussed above are general to twisted anisotropic homobilayers. In the following, we explore the moir\'e collapse from the perspective of \textit{ab initio} calculations including systems other than TBBP.

\section{Moir\'e Collapse as a Generic Feature of Twisted Anisotropic Homobilayers}

In this section, we summarize the \textit{ab initio} calculations. Here, we also present the results for twisted bilayer SnSe (TBSnSe) as an alternative anisotropic homobilayer exhibiting moir\'e collapse.
The first-principles structure relaxation and charge density calculations were carried out using VASP~\cite{Kohn1965, Kresse1996}. The projector-augmented wave pseudopotentials~\cite{{Blochl1994},{Kresse1999}} were used with a kinetic energy cutoff of 400~eV. The exchange-correlation (XC) functional was treated within the generalized gradient approximation of Perdew–Burke–Ernzerhof (PBE)~\cite{Perdew1996}.
The lattice constants of monolayer BP and SnSe were determined to be $l_x=3.294$~{\AA}, $l_y=4.565$~{\AA}, and  $l_x=4.331$~{\AA}, $l_y=4.271$~{\AA}, respectively.
The moir\'e supercell structures were constructed on the basis of the method described in Ref.~\cite{PhysRevB.96.041411}. The van der Waals interaction was handled within D3, with Becke-Johnson damping function~\cite{grimme2011effect} and non-local rev-vdW-DF2 methods, for TBBP and TBSnSe~\cite{revvdwdf2}, respectively. The bilayer structures were relaxed until the Helmann-Feynman force acting on every atom became smaller than 0.005~eV/$\textrm{\AA}$. Further, $\Gamma$-point, 1$\times$3$\times$1, 1$\times$4$\times$1 $k$-grid samples were used for the $90^\circ$ twisted supercells, near collapsed TBBP and TBSnSe, respectively. The electronic structure calculations were performed within the linear combination of atomic orbitals (LCAO) method implemented in QuantumATK~\cite{Smidstrup2019}. 

Figure~\ref{Fig3}(a) and (d) illustrate the possible commensurate supercells of TBBP and TBSnSe, respectively, highlight the number of atoms per moir\'e unit cell at various twist angles. The distribution of the coincidence-site lattices shown here is intricate, revealing a variety of large-angle structures with a wide range of atom counts per supercell. Notably, we emphasize that the two peaks nearest to the $90^{\circ}$ configurations occur at $90^\circ \pm \theta_M$, with high precision. To highlight this feature, we include two vertical lines at $90^\circ \pm \theta_M$ for both cases, where $\theta_M \approx 18.38^\circ$ for TBBP and $\theta_M \approx 0.8^\circ$ for TBSnSe. Furthermore, we confirm that the commensurate supercells around these regions are generally highly anisotropic~\cite{Snote}, indicating the tendency toward moir\'e collapse. While the secondary peaks may show a similar trend, they are outside the scope of this study and will be discussed in a separate work.

The electronic structure of fully relaxed moir\'e supercells are shown in Fig.~\ref{Fig3}(c)-(f). We selected the $90^\circ$ and nearly collapsed configurations for TBBP and TBSnSe. In both cases, the evolution of fully isotropic states at $90^\circ$ to highly anisotropic, quasi-1D, states at $\approx 90^\circ - \theta_M$ is clearly seen. While the TBBP results closely follow the continuum model described in this work, the electronic behavior in the TBSnSe system qualitatively deviates from the provided description due to the additional $p_x \pm ip_y$ character of the low-energy bands. Interestingly, TBSnSe transitions from an indirect band gap isotropic semiconductor to a direct band gap highly anisotropic semiconductor. Near the collapse, the direct energy gap is not localized at the $\Gamma$ point, suggesting the presence of quasi-1D valleys. This finding could open the door to the realization of Luttinger liquids, where additional valley degrees of freedom may play a significant role.

The quasi-1D behavior can be more clearly observed through the real-space charge distribution of the low-energy states. Figure~\ref{Fig3}(g) and (f) display the top and side view profiles of the charge density for TBBP and TBSnSe, respectively, corresponding to the lowest-energy conduction eigenstate at the $\Gamma$ point. The moir\'e unit cells are highlighted. We emphasize that these quasi-1D channels near the moir\'e collapse are highly dispersive due to the few angstroms long unit cell along $\textbf{a}_{1M}$, a lacking feature at small twist angles quasi-1D channels where the quasi-flat band in all directions tends to develop~\cite{PhysRevB.96.195406}.

\section{Perspectives}
We have shown here how twisted bilayers based on a variety of two-dimensional materials show quasi-one-dimensional behavior for "magic" twist angles near $\theta = 90^\circ$. The existence of this regime is based on the geometry of the layers. The lattices that we have studied need only to have orthorhombic symmetry, which is found in a number of families of 2D materials. 
\\ \indent
We studied the electronic properties of these systems using {\it ab initio} methods to extract continuum models of general validity. It is worth mentioning that the continuum models required to study these materials differ significantly from the continuum models used of twisted bilayer graphene and related systems. 
\\ \indent
The collapse of the moir\'e Brillouin zone gives rise to two or more incommensurate periodicities, opening the door to quasicrystalline behavior and fractal electronic spectra. Our results further reveal a rich variety of phenomena emerging near the collapse magic angle, particularly when electron-electron interactions are taken into account. Furthermore, the addition of more than two layers is expected to yield even more intricate and complex physical behaviors.

While a full classification of the topological properties of the electronic bands lies beyond the scope of this work, we anticipate that these features may play a significant role in shaping nonlinear transport and optical responses.

From a broader perspective, twisted anisotropic materials present an exciting frontier for engineering novel electronic phases, with moir\'e collapse potentially driving the emergence of a Luttinger liquid state. The Luttinger liquid phase is marked by collective excitations and non‐Fermi‐liquid behavior, providing a fertile playground to explore exotic pairing mechanisms and the interplay between topological and correlated physics~\cite{Wang2022, Yu2023, Jompol2009, PhysRevX.7.031009, PhysRevB.89.085101}. As experimental techniques for fabricating and probing twisted van der Waals structures improve, such systems could establish a new paradigm for designing low‐dimensional correlated electron phases, with potential applications ranging from quantum information platforms to tunable, strongly correlated electronics.

\section{Acknowledgements}
D. S, S. L and T. L acknowledge partial support from Office of Naval Research MURI grant N00014-23-1-2567. 
F. G. acknowledges support from
NOVMOMAT, project PID2022-142162NB-I00 funded by
MICIU/AEI/10.13039/501100011033 and by FEDER, UE as
well as financial support through the (MAD2D-CM)-MRR
MATERIALES AVANZADOS-IMDEA-NC.G.P.-M. is
supported by Comunidad de Madrid through the PIPF2022
programme (grant number PIPF-2022TEC-26326). This research
was supported in part by grant NSF PHY-2309135 to the
Kavli Institute for Theoretical Physics (KITP).

\section{Author Contributions}
D.S., F.G. and T.L. conceived the theory. D.S. developed the continuum model and performed the electronic structure analysis. S.L. performed the density functional theory calculation and structural analysis. D.S. wrote the manuscript. All authors contributed to the general scientific interpretation and edited the manuscript.

\section{Competing Interests}
The authors declare no competing interests.

\section{Additional Information}
\textbf{Supplementary information}. The online version contains supplementary material available at https://doi.org/xx.xxxx/.

%\begin{figure}[t]
%\centerline{\includegraphics[width = \linewidth]{figs/FigS.pdf}}
%\caption{.}
%\label{Fig1}
%\end{figure}

...

\bibliographystyle{apsrev}% your bst file here
\bibliography{my.bib,black_phosphorus_3} %your bib file here

\begin{thebibliography}{119}
\expandafter\ifx\csname natexlab\endcsname\relax\def\natexlab#1{#1}\fi
\expandafter\ifx\csname bibnamefont\endcsname\relax
  \def\bibnamefont#1{#1}\fi
\expandafter\ifx\csname bibfnamefont\endcsname\relax
  \def\bibfnamefont#1{#1}\fi
\expandafter\ifx\csname citenamefont\endcsname\relax
  \def\citenamefont#1{#1}\fi
\expandafter\ifx\csname url\endcsname\relax
  \def\url#1{\texttt{#1}}\fi
\expandafter\ifx\csname urlprefix\endcsname\relax\def\urlprefix{URL }\fi
\providecommand{\bibinfo}[2]{#2}
\providecommand{\eprint}[2][]{\url{#2}}

\bibitem[{\citenamefont{Carr et~al.}(2017)\citenamefont{Carr, Massatt, Fang, Cazeaux, Luskin, and Kaxiras}}]{carr_twistronics_2017}
\bibinfo{author}{\bibfnamefont{S.}~\bibnamefont{Carr}}, \bibinfo{author}{\bibfnamefont{D.}~\bibnamefont{Massatt}}, \bibinfo{author}{\bibfnamefont{S.}~\bibnamefont{Fang}}, \bibinfo{author}{\bibfnamefont{P.}~\bibnamefont{Cazeaux}}, \bibinfo{author}{\bibfnamefont{M.}~\bibnamefont{Luskin}}, \bibnamefont{and} \bibinfo{author}{\bibfnamefont{E.}~\bibnamefont{Kaxiras}}, \bibinfo{journal}{Physical Review B} \textbf{\bibinfo{volume}{95}} (\bibinfo{year}{2017}), ISSN \bibinfo{issn}{2469-9950, 2469-9969}, \bibinfo{note}{publisher: American Physical Society (APS)}, \urlprefix\url{https://link.aps.org/doi/10.1103/PhysRevB.95.075420}.

\bibitem[{\citenamefont{Bistritzer and MacDonald}(2011)}]{bistritzer_moire_2011}
\bibinfo{author}{\bibfnamefont{R.}~\bibnamefont{Bistritzer}} \bibnamefont{and} \bibinfo{author}{\bibfnamefont{A.~H.} \bibnamefont{MacDonald}}, \bibinfo{journal}{Proceedings of the National Academy of Sciences} \textbf{\bibinfo{volume}{108}}, \bibinfo{pages}{12233} (\bibinfo{year}{2011}), ISSN \bibinfo{issn}{0027-8424, 1091-6490}, \urlprefix\url{https://pnas.org/doi/full/10.1073/pnas.1108174108}.

\bibitem[{\citenamefont{Kennes et~al.}(2021)\citenamefont{Kennes, Claassen, Xian, Georges, Millis, Hone, Dean, Basov, Pasupathy, and Rubio}}]{Kennes2021}
\bibinfo{author}{\bibfnamefont{D.~M.} \bibnamefont{Kennes}}, \bibinfo{author}{\bibfnamefont{M.}~\bibnamefont{Claassen}}, \bibinfo{author}{\bibfnamefont{L.}~\bibnamefont{Xian}}, \bibinfo{author}{\bibfnamefont{A.}~\bibnamefont{Georges}}, \bibinfo{author}{\bibfnamefont{A.~J.} \bibnamefont{Millis}}, \bibinfo{author}{\bibfnamefont{J.}~\bibnamefont{Hone}}, \bibinfo{author}{\bibfnamefont{C.~R.} \bibnamefont{Dean}}, \bibinfo{author}{\bibfnamefont{D.~N.} \bibnamefont{Basov}}, \bibinfo{author}{\bibfnamefont{A.~N.} \bibnamefont{Pasupathy}}, \bibnamefont{and} \bibinfo{author}{\bibfnamefont{A.}~\bibnamefont{Rubio}}, \bibinfo{journal}{Nature Physics} \textbf{\bibinfo{volume}{17}}, \bibinfo{pages}{155–163} (\bibinfo{year}{2021}), ISSN \bibinfo{issn}{1745-2481}, \urlprefix\url{http://dx.doi.org/10.1038/s41567-020-01154-3}.

\bibitem[{\citenamefont{Lopes dos Santos et~al.}(2007)\citenamefont{Lopes dos Santos, Peres, and Castro Neto}}]{lopesdossantos_graphene_2007}
\bibinfo{author}{\bibfnamefont{J.~M.~B.} \bibnamefont{Lopes dos Santos}}, \bibinfo{author}{\bibfnamefont{N.~M.~R.} \bibnamefont{Peres}}, \bibnamefont{and} \bibinfo{author}{\bibfnamefont{A.~H.} \bibnamefont{Castro Neto}}, \bibinfo{journal}{Physical Review Letters} \textbf{\bibinfo{volume}{99}}, \bibinfo{pages}{256802} (\bibinfo{year}{2007}), ISSN \bibinfo{issn}{0031-9007, 1079-7114}, \urlprefix\url{https://link.aps.org/doi/10.1103/PhysRevLett.99.256802}.

\bibitem[{\citenamefont{Cao et~al.}(2018{\natexlab{a}})\citenamefont{Cao, Fatemi, Fang, Watanabe, Taniguchi, Kaxiras, and Jarillo-Herrero}}]{cao_unconventional_2018}
\bibinfo{author}{\bibfnamefont{Y.}~\bibnamefont{Cao}}, \bibinfo{author}{\bibfnamefont{V.}~\bibnamefont{Fatemi}}, \bibinfo{author}{\bibfnamefont{S.}~\bibnamefont{Fang}}, \bibinfo{author}{\bibfnamefont{K.}~\bibnamefont{Watanabe}}, \bibinfo{author}{\bibfnamefont{T.}~\bibnamefont{Taniguchi}}, \bibinfo{author}{\bibfnamefont{E.}~\bibnamefont{Kaxiras}}, \bibnamefont{and} \bibinfo{author}{\bibfnamefont{P.}~\bibnamefont{Jarillo-Herrero}}, \bibinfo{journal}{Nature} \textbf{\bibinfo{volume}{556}}, \bibinfo{pages}{43} (\bibinfo{year}{2018}{\natexlab{a}}), ISSN \bibinfo{issn}{0028-0836, 1476-4687}, \urlprefix\url{https://www.nature.com/articles/nature26160}.

\bibitem[{\citenamefont{Cao et~al.}(2018{\natexlab{b}})\citenamefont{Cao, Fatemi, Demir, Fang, Tomarken, Luo, Sanchez-Yamagishi, Watanabe, Taniguchi, Kaxiras et~al.}}]{cao_correlated_2018}
\bibinfo{author}{\bibfnamefont{Y.}~\bibnamefont{Cao}}, \bibinfo{author}{\bibfnamefont{V.}~\bibnamefont{Fatemi}}, \bibinfo{author}{\bibfnamefont{A.}~\bibnamefont{Demir}}, \bibinfo{author}{\bibfnamefont{S.}~\bibnamefont{Fang}}, \bibinfo{author}{\bibfnamefont{S.~L.} \bibnamefont{Tomarken}}, \bibinfo{author}{\bibfnamefont{J.~Y.} \bibnamefont{Luo}}, \bibinfo{author}{\bibfnamefont{J.~D.} \bibnamefont{Sanchez-Yamagishi}}, \bibinfo{author}{\bibfnamefont{K.}~\bibnamefont{Watanabe}}, \bibinfo{author}{\bibfnamefont{T.}~\bibnamefont{Taniguchi}}, \bibinfo{author}{\bibfnamefont{E.}~\bibnamefont{Kaxiras}}, \bibnamefont{et~al.}, \bibinfo{journal}{Nature} \textbf{\bibinfo{volume}{556}}, \bibinfo{pages}{80} (\bibinfo{year}{2018}{\natexlab{b}}), ISSN \bibinfo{issn}{0028-0836, 1476-4687}, \urlprefix\url{https://www.nature.com/articles/nature26154}.

\bibitem[{\citenamefont{Song et~al.}(2019)\citenamefont{Song, Wang, Shi, Li, Fang, and Bernevig}}]{PhysRevLett.123.036401}
\bibinfo{author}{\bibfnamefont{Z.}~\bibnamefont{Song}}, \bibinfo{author}{\bibfnamefont{Z.}~\bibnamefont{Wang}}, \bibinfo{author}{\bibfnamefont{W.}~\bibnamefont{Shi}}, \bibinfo{author}{\bibfnamefont{G.}~\bibnamefont{Li}}, \bibinfo{author}{\bibfnamefont{C.}~\bibnamefont{Fang}}, \bibnamefont{and} \bibinfo{author}{\bibfnamefont{B.~A.} \bibnamefont{Bernevig}}, \bibinfo{journal}{Phys. Rev. Lett.} \textbf{\bibinfo{volume}{123}}, \bibinfo{pages}{036401} (\bibinfo{year}{2019}), \urlprefix\url{https://link.aps.org/doi/10.1103/PhysRevLett.123.036401}.

\bibitem[{\citenamefont{Song and Bernevig}(2022)}]{PhysRevLett.129.047601}
\bibinfo{author}{\bibfnamefont{Z.-D.} \bibnamefont{Song}} \bibnamefont{and} \bibinfo{author}{\bibfnamefont{B.~A.} \bibnamefont{Bernevig}}, \bibinfo{journal}{Phys. Rev. Lett.} \textbf{\bibinfo{volume}{129}}, \bibinfo{pages}{047601} (\bibinfo{year}{2022}), \urlprefix\url{https://link.aps.org/doi/10.1103/PhysRevLett.129.047601}.

\bibitem[{\citenamefont{Trambly De~Laissardière et~al.}(2010)\citenamefont{Trambly De~Laissardière, Mayou, and Magaud}}]{trambly_de_laissardiere_localization_2010}
\bibinfo{author}{\bibfnamefont{G.}~\bibnamefont{Trambly De~Laissardière}}, \bibinfo{author}{\bibfnamefont{D.}~\bibnamefont{Mayou}}, \bibnamefont{and} \bibinfo{author}{\bibfnamefont{L.}~\bibnamefont{Magaud}}, \bibinfo{journal}{Nano Letters} \textbf{\bibinfo{volume}{10}}, \bibinfo{pages}{804} (\bibinfo{year}{2010}), ISSN \bibinfo{issn}{1530-6984, 1530-6992}, \urlprefix\url{https://pubs.acs.org/doi/10.1021/nl902948m}.

\bibitem[{\citenamefont{Kapfer et~al.}(2023)\citenamefont{Kapfer, Jessen, Eisele, Fu, Danielsen, Darlington, Moore, Finney, Marchese, Hsieh et~al.}}]{kapfer_programming_2023}
\bibinfo{author}{\bibfnamefont{M.}~\bibnamefont{Kapfer}}, \bibinfo{author}{\bibfnamefont{B.~S.} \bibnamefont{Jessen}}, \bibinfo{author}{\bibfnamefont{M.~E.} \bibnamefont{Eisele}}, \bibinfo{author}{\bibfnamefont{M.}~\bibnamefont{Fu}}, \bibinfo{author}{\bibfnamefont{D.~R.} \bibnamefont{Danielsen}}, \bibinfo{author}{\bibfnamefont{T.~P.} \bibnamefont{Darlington}}, \bibinfo{author}{\bibfnamefont{S.~L.} \bibnamefont{Moore}}, \bibinfo{author}{\bibfnamefont{N.~R.} \bibnamefont{Finney}}, \bibinfo{author}{\bibfnamefont{A.}~\bibnamefont{Marchese}}, \bibinfo{author}{\bibfnamefont{V.}~\bibnamefont{Hsieh}}, \bibnamefont{et~al.}, \bibinfo{journal}{Science} \textbf{\bibinfo{volume}{381}}, \bibinfo{pages}{677} (\bibinfo{year}{2023}), ISSN \bibinfo{issn}{0036-8075, 1095-9203}, \urlprefix\url{https://www.science.org/doi/10.1126/science.ade9995}.

\bibitem[{\citenamefont{Peña et~al.}(2023)\citenamefont{Peña, Dey, Chowdhury, Azizimanesh, Hou, Sewaket, Watson, Askari, and Wu}}]{pena_moire_2023}
\bibinfo{author}{\bibfnamefont{T.}~\bibnamefont{Peña}}, \bibinfo{author}{\bibfnamefont{A.}~\bibnamefont{Dey}}, \bibinfo{author}{\bibfnamefont{S.~A.} \bibnamefont{Chowdhury}}, \bibinfo{author}{\bibfnamefont{A.}~\bibnamefont{Azizimanesh}}, \bibinfo{author}{\bibfnamefont{W.}~\bibnamefont{Hou}}, \bibinfo{author}{\bibfnamefont{A.}~\bibnamefont{Sewaket}}, \bibinfo{author}{\bibfnamefont{C.}~\bibnamefont{Watson}}, \bibinfo{author}{\bibfnamefont{H.}~\bibnamefont{Askari}}, \bibnamefont{and} \bibinfo{author}{\bibfnamefont{S.~M.} \bibnamefont{Wu}}, \bibinfo{journal}{Applied Physics Letters} \textbf{\bibinfo{volume}{122}}, \bibinfo{pages}{143101} (\bibinfo{year}{2023}), ISSN \bibinfo{issn}{0003-6951, 1077-3118}, \urlprefix\url{https://pubs.aip.org/apl/article/122/14/143101/2882378/Moire-engineering-in-2D-heterostructures-with}.

\bibitem[{\citenamefont{Angeli and MacDonald}(2021)}]{angeli__2021}
\bibinfo{author}{\bibfnamefont{M.}~\bibnamefont{Angeli}} \bibnamefont{and} \bibinfo{author}{\bibfnamefont{A.~H.} \bibnamefont{MacDonald}}, \bibinfo{journal}{Proceedings of the National Academy of Sciences} \textbf{\bibinfo{volume}{118}}, \bibinfo{pages}{e2021826118} (\bibinfo{year}{2021}), ISSN \bibinfo{issn}{0027-8424, 1091-6490}, \urlprefix\url{https://pnas.org/doi/full/10.1073/pnas.2021826118}.

\bibitem[{\citenamefont{Guinea and Walet}(2019)}]{guinea_continuum_2019}
\bibinfo{author}{\bibfnamefont{F.}~\bibnamefont{Guinea}} \bibnamefont{and} \bibinfo{author}{\bibfnamefont{N.~R.} \bibnamefont{Walet}}, \bibinfo{journal}{Physical Review B} \textbf{\bibinfo{volume}{99}}, \bibinfo{pages}{205134} (\bibinfo{year}{2019}), ISSN \bibinfo{issn}{2469-9950, 2469-9969}, \urlprefix\url{https://link.aps.org/doi/10.1103/PhysRevB.99.205134}.

\bibitem[{\citenamefont{Naik and Jain}(2018)}]{naik_ultraflatbands_2018}
\bibinfo{author}{\bibfnamefont{M.~H.} \bibnamefont{Naik}} \bibnamefont{and} \bibinfo{author}{\bibfnamefont{M.}~\bibnamefont{Jain}}, \bibinfo{journal}{Physical Review Letters} \textbf{\bibinfo{volume}{121}}, \bibinfo{pages}{266401} (\bibinfo{year}{2018}), ISSN \bibinfo{issn}{0031-9007, 1079-7114}, \urlprefix\url{https://link.aps.org/doi/10.1103/PhysRevLett.121.266401}.

\bibitem[{\citenamefont{Huder et~al.}(2018)\citenamefont{Huder, Artaud, Le~Quang, De~Laissardière, Jansen, Lapertot, Chapelier, and Renard}}]{huder_electronic_2018}
\bibinfo{author}{\bibfnamefont{L.}~\bibnamefont{Huder}}, \bibinfo{author}{\bibfnamefont{A.}~\bibnamefont{Artaud}}, \bibinfo{author}{\bibfnamefont{T.}~\bibnamefont{Le~Quang}}, \bibinfo{author}{\bibfnamefont{G.~T.} \bibnamefont{De~Laissardière}}, \bibinfo{author}{\bibfnamefont{A.~G.} \bibnamefont{Jansen}}, \bibinfo{author}{\bibfnamefont{G.}~\bibnamefont{Lapertot}}, \bibinfo{author}{\bibfnamefont{C.}~\bibnamefont{Chapelier}}, \bibnamefont{and} \bibinfo{author}{\bibfnamefont{V.~T.} \bibnamefont{Renard}}, \bibinfo{journal}{Physical Review Letters} \textbf{\bibinfo{volume}{120}}, \bibinfo{pages}{156405} (\bibinfo{year}{2018}), ISSN \bibinfo{issn}{0031-9007, 1079-7114}, \urlprefix\url{https://link.aps.org/doi/10.1103/PhysRevLett.120.156405}.

\bibitem[{\citenamefont{Lopes Dos~Santos et~al.}(2012)\citenamefont{Lopes Dos~Santos, Peres, and Castro~Neto}}]{lopes_dos_santos_continuum_2012}
\bibinfo{author}{\bibfnamefont{J.~M.~B.} \bibnamefont{Lopes Dos~Santos}}, \bibinfo{author}{\bibfnamefont{N.~M.~R.} \bibnamefont{Peres}}, \bibnamefont{and} \bibinfo{author}{\bibfnamefont{A.~H.} \bibnamefont{Castro~Neto}}, \bibinfo{journal}{Physical Review B} \textbf{\bibinfo{volume}{86}}, \bibinfo{pages}{155449} (\bibinfo{year}{2012}), ISSN \bibinfo{issn}{1098-0121, 1550-235X}, \urlprefix\url{https://link.aps.org/doi/10.1103/PhysRevB.86.155449}.

\bibitem[{\citenamefont{Li et~al.}(2024)\citenamefont{Li, Xu, Wang, Han, Watanabe, Taniguchi, Song, Ma, Gao, Jiang et~al.}}]{li_quasiperiodic_2024}
\bibinfo{author}{\bibfnamefont{S.-y.} \bibnamefont{Li}}, \bibinfo{author}{\bibfnamefont{Z.}~\bibnamefont{Xu}}, \bibinfo{author}{\bibfnamefont{Y.}~\bibnamefont{Wang}}, \bibinfo{author}{\bibfnamefont{Y.}~\bibnamefont{Han}}, \bibinfo{author}{\bibfnamefont{K.}~\bibnamefont{Watanabe}}, \bibinfo{author}{\bibfnamefont{T.}~\bibnamefont{Taniguchi}}, \bibinfo{author}{\bibfnamefont{A.}~\bibnamefont{Song}}, \bibinfo{author}{\bibfnamefont{T.-B.} \bibnamefont{Ma}}, \bibinfo{author}{\bibfnamefont{H.-J.} \bibnamefont{Gao}}, \bibinfo{author}{\bibfnamefont{Y.}~\bibnamefont{Jiang}}, \bibnamefont{et~al.}, \bibinfo{journal}{Physical Review Letters} \textbf{\bibinfo{volume}{133}}, \bibinfo{pages}{196401} (\bibinfo{year}{2024}), ISSN \bibinfo{issn}{0031-9007, 1079-7114}, \urlprefix\url{https://link.aps.org/doi/10.1103/PhysRevLett.133.196401}.

\bibitem[{\citenamefont{Lee et~al.}(2024)\citenamefont{Lee, de~Sousa, Jalan, and Low}}]{Lee2024}
\bibinfo{author}{\bibfnamefont{S.}~\bibnamefont{Lee}}, \bibinfo{author}{\bibfnamefont{D.~J.~P.} \bibnamefont{de~Sousa}}, \bibinfo{author}{\bibfnamefont{B.}~\bibnamefont{Jalan}}, \bibnamefont{and} \bibinfo{author}{\bibfnamefont{T.}~\bibnamefont{Low}}, \bibinfo{journal}{Science Advances} \textbf{\bibinfo{volume}{10}} (\bibinfo{year}{2024}), ISSN \bibinfo{issn}{2375-2548}, \urlprefix\url{http://dx.doi.org/10.1126/sciadv.adq0293}.

\bibitem[{\citenamefont{Larson et~al.}(2024)\citenamefont{Larson, Bennett, Ali, Chaves, Arora, Rabe, and Kaxiras}}]{https://doi.org/10.48550/arxiv.2411.16497}
\bibinfo{author}{\bibfnamefont{D.~T.} \bibnamefont{Larson}}, \bibinfo{author}{\bibfnamefont{D.}~\bibnamefont{Bennett}}, \bibinfo{author}{\bibfnamefont{A.}~\bibnamefont{Ali}}, \bibinfo{author}{\bibfnamefont{A.~S.} \bibnamefont{Chaves}}, \bibinfo{author}{\bibfnamefont{R.}~\bibnamefont{Arora}}, \bibinfo{author}{\bibfnamefont{K.~M.} \bibnamefont{Rabe}}, \bibnamefont{and} \bibinfo{author}{\bibfnamefont{E.}~\bibnamefont{Kaxiras}}, \emph{\bibinfo{title}{Stacking-dependent electronic structure of ultrathin perovskite bilayers}} (\bibinfo{year}{2024}), \urlprefix\url{https://arxiv.org/abs/2411.16497}.

\bibitem[{\citenamefont{de~Sousa et~al.}(2024{\natexlab{a}})\citenamefont{de~Sousa, Lee, and Low}}]{https://doi.org/10.48550/arxiv.2409.06806}
\bibinfo{author}{\bibfnamefont{D.~J.~P.} \bibnamefont{de~Sousa}}, \bibinfo{author}{\bibfnamefont{S.}~\bibnamefont{Lee}}, \bibnamefont{and} \bibinfo{author}{\bibfnamefont{T.}~\bibnamefont{Low}}, \emph{\bibinfo{title}{Moiré kramers-weyl fermions from structural chirality with ideal radial spin texture}} (\bibinfo{year}{2024}{\natexlab{a}}), \urlprefix\url{https://arxiv.org/abs/2409.06806}.

\bibitem[{\citenamefont{Slot et~al.}(2023)\citenamefont{Slot, Maximenko, Haney, Kim, Walkup, Strelcov, Le, Shih, Yildiz, Blankenship et~al.}}]{Slot2023}
\bibinfo{author}{\bibfnamefont{M.~R.} \bibnamefont{Slot}}, \bibinfo{author}{\bibfnamefont{Y.}~\bibnamefont{Maximenko}}, \bibinfo{author}{\bibfnamefont{P.~M.} \bibnamefont{Haney}}, \bibinfo{author}{\bibfnamefont{S.}~\bibnamefont{Kim}}, \bibinfo{author}{\bibfnamefont{D.~T.} \bibnamefont{Walkup}}, \bibinfo{author}{\bibfnamefont{E.}~\bibnamefont{Strelcov}}, \bibinfo{author}{\bibfnamefont{S.~T.} \bibnamefont{Le}}, \bibinfo{author}{\bibfnamefont{E.~M.} \bibnamefont{Shih}}, \bibinfo{author}{\bibfnamefont{D.}~\bibnamefont{Yildiz}}, \bibinfo{author}{\bibfnamefont{S.~R.} \bibnamefont{Blankenship}}, \bibnamefont{et~al.}, \bibinfo{journal}{Science} \textbf{\bibinfo{volume}{382}}, \bibinfo{pages}{81–87} (\bibinfo{year}{2023}), ISSN \bibinfo{issn}{1095-9203}, \urlprefix\url{http://dx.doi.org/10.1126/science.adf2040}.

\bibitem[{\citenamefont{Yoo et~al.}(2019)\citenamefont{Yoo, Engelke, Carr, Fang, Zhang, Cazeaux, Sung, Hovden, Tsen, Taniguchi et~al.}}]{Yoo2019}
\bibinfo{author}{\bibfnamefont{H.}~\bibnamefont{Yoo}}, \bibinfo{author}{\bibfnamefont{R.}~\bibnamefont{Engelke}}, \bibinfo{author}{\bibfnamefont{S.}~\bibnamefont{Carr}}, \bibinfo{author}{\bibfnamefont{S.}~\bibnamefont{Fang}}, \bibinfo{author}{\bibfnamefont{K.}~\bibnamefont{Zhang}}, \bibinfo{author}{\bibfnamefont{P.}~\bibnamefont{Cazeaux}}, \bibinfo{author}{\bibfnamefont{S.~H.} \bibnamefont{Sung}}, \bibinfo{author}{\bibfnamefont{R.}~\bibnamefont{Hovden}}, \bibinfo{author}{\bibfnamefont{A.~W.} \bibnamefont{Tsen}}, \bibinfo{author}{\bibfnamefont{T.}~\bibnamefont{Taniguchi}}, \bibnamefont{et~al.}, \bibinfo{journal}{Nature Materials} \textbf{\bibinfo{volume}{18}}, \bibinfo{pages}{448–453} (\bibinfo{year}{2019}), ISSN \bibinfo{issn}{1476-4660}, \urlprefix\url{http://dx.doi.org/10.1038/s41563-019-0346-z}.

\bibitem[{\citenamefont{Wu et~al.}(2018)\citenamefont{Wu, Lovorn, Tutuc, and MacDonald}}]{PhysRevLett.121.026402}
\bibinfo{author}{\bibfnamefont{F.}~\bibnamefont{Wu}}, \bibinfo{author}{\bibfnamefont{T.}~\bibnamefont{Lovorn}}, \bibinfo{author}{\bibfnamefont{E.}~\bibnamefont{Tutuc}}, \bibnamefont{and} \bibinfo{author}{\bibfnamefont{A.~H.} \bibnamefont{MacDonald}}, \bibinfo{journal}{Phys. Rev. Lett.} \textbf{\bibinfo{volume}{121}}, \bibinfo{pages}{026402} (\bibinfo{year}{2018}), \urlprefix\url{https://link.aps.org/doi/10.1103/PhysRevLett.121.026402}.

\bibitem[{\citenamefont{San-Jose et~al.}(2012)\citenamefont{San-Jose, González, and Guinea}}]{san-jose_non-abelian_2012}
\bibinfo{author}{\bibfnamefont{P.}~\bibnamefont{San-Jose}}, \bibinfo{author}{\bibfnamefont{J.}~\bibnamefont{González}}, \bibnamefont{and} \bibinfo{author}{\bibfnamefont{F.}~\bibnamefont{Guinea}}, \bibinfo{journal}{Physical Review Letters} \textbf{\bibinfo{volume}{108}}, \bibinfo{pages}{216802} (\bibinfo{year}{2012}), ISSN \bibinfo{issn}{0031-9007, 1079-7114}, \urlprefix\url{https://link.aps.org/doi/10.1103/PhysRevLett.108.216802}.

\bibitem[{\citenamefont{Devakul et~al.}(2021)\citenamefont{Devakul, Crépel, Zhang, and Fu}}]{Devakul2021}
\bibinfo{author}{\bibfnamefont{T.}~\bibnamefont{Devakul}}, \bibinfo{author}{\bibfnamefont{V.}~\bibnamefont{Crépel}}, \bibinfo{author}{\bibfnamefont{Y.}~\bibnamefont{Zhang}}, \bibnamefont{and} \bibinfo{author}{\bibfnamefont{L.}~\bibnamefont{Fu}}, \bibinfo{journal}{Nature Communications} \textbf{\bibinfo{volume}{12}} (\bibinfo{year}{2021}), ISSN \bibinfo{issn}{2041-1723}, \urlprefix\url{http://dx.doi.org/10.1038/s41467-021-27042-9}.

\bibitem[{\citenamefont{Wang et~al.}(2022)\citenamefont{Wang, Yu, Kwan, Jia, Lei, Klemenz, Cevallos, Singha, Devakul, Watanabe et~al.}}]{Wang2022}
\bibinfo{author}{\bibfnamefont{P.}~\bibnamefont{Wang}}, \bibinfo{author}{\bibfnamefont{G.}~\bibnamefont{Yu}}, \bibinfo{author}{\bibfnamefont{Y.~H.} \bibnamefont{Kwan}}, \bibinfo{author}{\bibfnamefont{Y.}~\bibnamefont{Jia}}, \bibinfo{author}{\bibfnamefont{S.}~\bibnamefont{Lei}}, \bibinfo{author}{\bibfnamefont{S.}~\bibnamefont{Klemenz}}, \bibinfo{author}{\bibfnamefont{F.~A.} \bibnamefont{Cevallos}}, \bibinfo{author}{\bibfnamefont{R.}~\bibnamefont{Singha}}, \bibinfo{author}{\bibfnamefont{T.}~\bibnamefont{Devakul}}, \bibinfo{author}{\bibfnamefont{K.}~\bibnamefont{Watanabe}}, \bibnamefont{et~al.}, \bibinfo{journal}{Nature} \textbf{\bibinfo{volume}{605}}, \bibinfo{pages}{57–62} (\bibinfo{year}{2022}), ISSN \bibinfo{issn}{1476-4687}, \urlprefix\url{http://dx.doi.org/10.1038/s41586-022-04514-6}.

\bibitem[{\citenamefont{Yu et~al.}(2023)\citenamefont{Yu, Wang, Uzan-Narovlansky, Jia, Onyszczak, Singha, Gui, Song, Tang, Watanabe et~al.}}]{Yu2023}
\bibinfo{author}{\bibfnamefont{G.}~\bibnamefont{Yu}}, \bibinfo{author}{\bibfnamefont{P.}~\bibnamefont{Wang}}, \bibinfo{author}{\bibfnamefont{A.~J.} \bibnamefont{Uzan-Narovlansky}}, \bibinfo{author}{\bibfnamefont{Y.}~\bibnamefont{Jia}}, \bibinfo{author}{\bibfnamefont{M.}~\bibnamefont{Onyszczak}}, \bibinfo{author}{\bibfnamefont{R.}~\bibnamefont{Singha}}, \bibinfo{author}{\bibfnamefont{X.}~\bibnamefont{Gui}}, \bibinfo{author}{\bibfnamefont{T.}~\bibnamefont{Song}}, \bibinfo{author}{\bibfnamefont{Y.}~\bibnamefont{Tang}}, \bibinfo{author}{\bibfnamefont{K.}~\bibnamefont{Watanabe}}, \bibnamefont{et~al.}, \bibinfo{journal}{Nature Communications} \textbf{\bibinfo{volume}{14}} (\bibinfo{year}{2023}), ISSN \bibinfo{issn}{2041-1723}, \urlprefix\url{http://dx.doi.org/10.1038/s41467-023-42821-2}.

\bibitem[{\citenamefont{Sharpe et~al.}(2019)\citenamefont{Sharpe, Fox, Barnard, Finney, Watanabe, Taniguchi, Kastner, and Goldhaber-Gordon}}]{Sharpe2019}
\bibinfo{author}{\bibfnamefont{A.~L.} \bibnamefont{Sharpe}}, \bibinfo{author}{\bibfnamefont{E.~J.} \bibnamefont{Fox}}, \bibinfo{author}{\bibfnamefont{A.~W.} \bibnamefont{Barnard}}, \bibinfo{author}{\bibfnamefont{J.}~\bibnamefont{Finney}}, \bibinfo{author}{\bibfnamefont{K.}~\bibnamefont{Watanabe}}, \bibinfo{author}{\bibfnamefont{T.}~\bibnamefont{Taniguchi}}, \bibinfo{author}{\bibfnamefont{M.~A.} \bibnamefont{Kastner}}, \bibnamefont{and} \bibinfo{author}{\bibfnamefont{D.}~\bibnamefont{Goldhaber-Gordon}}, \bibinfo{journal}{Science} \textbf{\bibinfo{volume}{365}}, \bibinfo{pages}{605–608} (\bibinfo{year}{2019}), ISSN \bibinfo{issn}{1095-9203}, \urlprefix\url{http://dx.doi.org/10.1126/science.aaw3780}.

\bibitem[{\citenamefont{Serlin et~al.}(2020)\citenamefont{Serlin, Tschirhart, Polshyn, Zhang, Zhu, Watanabe, Taniguchi, Balents, and Young}}]{Serlin2020}
\bibinfo{author}{\bibfnamefont{M.}~\bibnamefont{Serlin}}, \bibinfo{author}{\bibfnamefont{C.~L.} \bibnamefont{Tschirhart}}, \bibinfo{author}{\bibfnamefont{H.}~\bibnamefont{Polshyn}}, \bibinfo{author}{\bibfnamefont{Y.}~\bibnamefont{Zhang}}, \bibinfo{author}{\bibfnamefont{J.}~\bibnamefont{Zhu}}, \bibinfo{author}{\bibfnamefont{K.}~\bibnamefont{Watanabe}}, \bibinfo{author}{\bibfnamefont{T.}~\bibnamefont{Taniguchi}}, \bibinfo{author}{\bibfnamefont{L.}~\bibnamefont{Balents}}, \bibnamefont{and} \bibinfo{author}{\bibfnamefont{A.~F.} \bibnamefont{Young}}, \bibinfo{journal}{Science} \textbf{\bibinfo{volume}{367}}, \bibinfo{pages}{900–903} (\bibinfo{year}{2020}), ISSN \bibinfo{issn}{1095-9203}, \urlprefix\url{http://dx.doi.org/10.1126/science.aay5533}.

\bibitem[{\citenamefont{Rudenko and Katsnelson}(2024)}]{rudenko_anisotropic_2024}
\bibinfo{author}{\bibfnamefont{A.~N.} \bibnamefont{Rudenko}} \bibnamefont{and} \bibinfo{author}{\bibfnamefont{M.~I.} \bibnamefont{Katsnelson}}, \bibinfo{journal}{2D Materials} \textbf{\bibinfo{volume}{11}}, \bibinfo{pages}{042002} (\bibinfo{year}{2024}), ISSN \bibinfo{issn}{2053-1583}, \bibinfo{note}{publisher: IOP Publishing}, \urlprefix\url{https://iopscience.iop.org/article/10.1088/2053-1583/ad64e1}.

\bibitem[{\citenamefont{Li et~al.}(2019)\citenamefont{Li, Han, Pi, Niu, Han, Wang, Su, Li, Xiong, Bando et~al.}}]{Li2019}
\bibinfo{author}{\bibfnamefont{L.}~\bibnamefont{Li}}, \bibinfo{author}{\bibfnamefont{W.}~\bibnamefont{Han}}, \bibinfo{author}{\bibfnamefont{L.}~\bibnamefont{Pi}}, \bibinfo{author}{\bibfnamefont{P.}~\bibnamefont{Niu}}, \bibinfo{author}{\bibfnamefont{J.}~\bibnamefont{Han}}, \bibinfo{author}{\bibfnamefont{C.}~\bibnamefont{Wang}}, \bibinfo{author}{\bibfnamefont{B.}~\bibnamefont{Su}}, \bibinfo{author}{\bibfnamefont{H.}~\bibnamefont{Li}}, \bibinfo{author}{\bibfnamefont{J.}~\bibnamefont{Xiong}}, \bibinfo{author}{\bibfnamefont{Y.}~\bibnamefont{Bando}}, \bibnamefont{et~al.}, \bibinfo{journal}{InfoMat} \textbf{\bibinfo{volume}{1}}, \bibinfo{pages}{54–73} (\bibinfo{year}{2019}), ISSN \bibinfo{issn}{2567-3165}, \urlprefix\url{http://dx.doi.org/10.1002/inf2.12005}.

\bibitem[{\citenamefont{Lin et~al.}(2016)\citenamefont{Lin, Grassi, Low, and Helmy}}]{lin_multilayer_2016}
\bibinfo{author}{\bibfnamefont{C.}~\bibnamefont{Lin}}, \bibinfo{author}{\bibfnamefont{R.}~\bibnamefont{Grassi}}, \bibinfo{author}{\bibfnamefont{T.}~\bibnamefont{Low}}, \bibnamefont{and} \bibinfo{author}{\bibfnamefont{A.~S.} \bibnamefont{Helmy}}, \bibinfo{journal}{Nano Letters} \textbf{\bibinfo{volume}{16}}, \bibinfo{pages}{1683} (\bibinfo{year}{2016}), ISSN \bibinfo{issn}{1530-6984, 1530-6992}, \bibinfo{note}{publisher: American Chemical Society (ACS)}, \urlprefix\url{https://pubs.acs.org/doi/10.1021/acs.nanolett.5b04594}.

\bibitem[{\citenamefont{Wu et~al.}(2015)\citenamefont{Wu, Topsakal, Low, Robbins, Haratipour, Jeong, Wentzcovitch, Koester, and Mkhoyan}}]{wu_atomic_2015}
\bibinfo{author}{\bibfnamefont{R.~J.} \bibnamefont{Wu}}, \bibinfo{author}{\bibfnamefont{M.}~\bibnamefont{Topsakal}}, \bibinfo{author}{\bibfnamefont{T.}~\bibnamefont{Low}}, \bibinfo{author}{\bibfnamefont{M.~C.} \bibnamefont{Robbins}}, \bibinfo{author}{\bibfnamefont{N.}~\bibnamefont{Haratipour}}, \bibinfo{author}{\bibfnamefont{J.~S.} \bibnamefont{Jeong}}, \bibinfo{author}{\bibfnamefont{R.~M.} \bibnamefont{Wentzcovitch}}, \bibinfo{author}{\bibfnamefont{S.~J.} \bibnamefont{Koester}}, \bibnamefont{and} \bibinfo{author}{\bibfnamefont{K.~A.} \bibnamefont{Mkhoyan}}, \bibinfo{journal}{Journal of Vacuum Science \& Technology A: Vacuum, Surfaces, and Films} \textbf{\bibinfo{volume}{33}} (\bibinfo{year}{2015}), ISSN \bibinfo{issn}{0734-2101, 1520-8559}, \bibinfo{note}{publisher: American Vacuum Society}, \urlprefix\url{https://pubs.aip.org/jva/article/33/6/060604/246797/Atomic-and-electronic-structure-of-exfoliated}.

\bibitem[{\citenamefont{Rudenko et~al.}(2016)\citenamefont{Rudenko, Brener, and Katsnelson}}]{rudenko_intrinsic_2016}
\bibinfo{author}{\bibfnamefont{A.}~\bibnamefont{Rudenko}}, \bibinfo{author}{\bibfnamefont{S.}~\bibnamefont{Brener}}, \bibnamefont{and} \bibinfo{author}{\bibfnamefont{M.}~\bibnamefont{Katsnelson}}, \bibinfo{journal}{Physical Review Letters} \textbf{\bibinfo{volume}{116}} (\bibinfo{year}{2016}), ISSN \bibinfo{issn}{0031-9007, 1079-7114}, \bibinfo{note}{publisher: American Physical Society (APS)}, \urlprefix\url{https://link.aps.org/doi/10.1103/PhysRevLett.116.246401}.

\bibitem[{\citenamefont{Liu et~al.}(2016)\citenamefont{Liu, Low, and Ruden}}]{liu_mobility_2016}
\bibinfo{author}{\bibfnamefont{Y.}~\bibnamefont{Liu}}, \bibinfo{author}{\bibfnamefont{T.}~\bibnamefont{Low}}, \bibnamefont{and} \bibinfo{author}{\bibfnamefont{P.~P.} \bibnamefont{Ruden}}, \bibinfo{journal}{Physical Review B} \textbf{\bibinfo{volume}{93}} (\bibinfo{year}{2016}), ISSN \bibinfo{issn}{2469-9950, 2469-9969}, \bibinfo{note}{publisher: American Physical Society (APS)}, \urlprefix\url{https://link.aps.org/doi/10.1103/PhysRevB.93.165402}.

\bibitem[{\citenamefont{Ling et~al.}(2015)\citenamefont{Ling, Wang, Huang, Xia, and Dresselhaus}}]{ling_renaissance_2015}
\bibinfo{author}{\bibfnamefont{X.}~\bibnamefont{Ling}}, \bibinfo{author}{\bibfnamefont{H.}~\bibnamefont{Wang}}, \bibinfo{author}{\bibfnamefont{S.}~\bibnamefont{Huang}}, \bibinfo{author}{\bibfnamefont{F.}~\bibnamefont{Xia}}, \bibnamefont{and} \bibinfo{author}{\bibfnamefont{M.~S.} \bibnamefont{Dresselhaus}}, \bibinfo{journal}{Proceedings of the National Academy of Sciences} \textbf{\bibinfo{volume}{112}}, \bibinfo{pages}{4523} (\bibinfo{year}{2015}), ISSN \bibinfo{issn}{0027-8424, 1091-6490}, \bibinfo{note}{publisher: Proceedings of the National Academy of Sciences}, \urlprefix\url{https://pnas.org/doi/full/10.1073/pnas.1416581112}.

\bibitem[{\citenamefont{Castellanos-Gomez et~al.}(2014)\citenamefont{Castellanos-Gomez, Vicarelli, Prada, Island, Narasimha-Acharya, Blanter, Groenendijk, Buscema, Steele, Alvarez et~al.}}]{castellanos-gomez_isolation_2014}
\bibinfo{author}{\bibfnamefont{A.}~\bibnamefont{Castellanos-Gomez}}, \bibinfo{author}{\bibfnamefont{L.}~\bibnamefont{Vicarelli}}, \bibinfo{author}{\bibfnamefont{E.}~\bibnamefont{Prada}}, \bibinfo{author}{\bibfnamefont{J.~O.} \bibnamefont{Island}}, \bibinfo{author}{\bibfnamefont{K.~L.} \bibnamefont{Narasimha-Acharya}}, \bibinfo{author}{\bibfnamefont{S.~I.} \bibnamefont{Blanter}}, \bibinfo{author}{\bibfnamefont{D.~J.} \bibnamefont{Groenendijk}}, \bibinfo{author}{\bibfnamefont{M.}~\bibnamefont{Buscema}}, \bibinfo{author}{\bibfnamefont{G.~A.} \bibnamefont{Steele}}, \bibinfo{author}{\bibfnamefont{J.~V.} \bibnamefont{Alvarez}}, \bibnamefont{et~al.}, \bibinfo{journal}{2D Materials} \textbf{\bibinfo{volume}{1}}, \bibinfo{pages}{025001} (\bibinfo{year}{2014}), ISSN \bibinfo{issn}{2053-1583}, \bibinfo{note}{publisher: IOP Publishing}, \urlprefix\url{https://iopscience.iop.org/article/10.1088/2053-1583/1/2/025001}.

\bibitem[{\citenamefont{Xia et~al.}(2014)\citenamefont{Xia, Wang, and Jia}}]{xia_rediscovering_2014}
\bibinfo{author}{\bibfnamefont{F.}~\bibnamefont{Xia}}, \bibinfo{author}{\bibfnamefont{H.}~\bibnamefont{Wang}}, \bibnamefont{and} \bibinfo{author}{\bibfnamefont{Y.}~\bibnamefont{Jia}}, \bibinfo{journal}{Nature Communications} \textbf{\bibinfo{volume}{5}} (\bibinfo{year}{2014}), ISSN \bibinfo{issn}{2041-1723}, \bibinfo{note}{publisher: Springer Science and Business Media LLC}, \urlprefix\url{https://www.nature.com/articles/ncomms5458}.

\bibitem[{\citenamefont{Tran et~al.}(2014)\citenamefont{Tran, Soklaski, Liang, and Yang}}]{tran_layer-controlled_2014}
\bibinfo{author}{\bibfnamefont{V.}~\bibnamefont{Tran}}, \bibinfo{author}{\bibfnamefont{R.}~\bibnamefont{Soklaski}}, \bibinfo{author}{\bibfnamefont{Y.}~\bibnamefont{Liang}}, \bibnamefont{and} \bibinfo{author}{\bibfnamefont{L.}~\bibnamefont{Yang}}, \bibinfo{journal}{Physical Review B} \textbf{\bibinfo{volume}{89}} (\bibinfo{year}{2014}), ISSN \bibinfo{issn}{1098-0121, 1550-235X}, \bibinfo{note}{publisher: American Physical Society (APS)}, \urlprefix\url{https://link.aps.org/doi/10.1103/PhysRevB.89.235319}.

\bibitem[{\citenamefont{Liu et~al.}(2014)\citenamefont{Liu, Neal, Zhu, Luo, Xu, Tománek, and Ye}}]{liu_phosphorene_2014}
\bibinfo{author}{\bibfnamefont{H.}~\bibnamefont{Liu}}, \bibinfo{author}{\bibfnamefont{A.~T.} \bibnamefont{Neal}}, \bibinfo{author}{\bibfnamefont{Z.}~\bibnamefont{Zhu}}, \bibinfo{author}{\bibfnamefont{Z.}~\bibnamefont{Luo}}, \bibinfo{author}{\bibfnamefont{X.}~\bibnamefont{Xu}}, \bibinfo{author}{\bibfnamefont{D.}~\bibnamefont{Tománek}}, \bibnamefont{and} \bibinfo{author}{\bibfnamefont{P.~D.} \bibnamefont{Ye}}, \bibinfo{journal}{ACS Nano} \textbf{\bibinfo{volume}{8}}, \bibinfo{pages}{4033} (\bibinfo{year}{2014}), ISSN \bibinfo{issn}{1936-0851, 1936-086X}, \bibinfo{note}{publisher: American Chemical Society (ACS)}, \urlprefix\url{https://pubs.acs.org/doi/10.1021/nn501226z}.

\bibitem[{\citenamefont{Koenig et~al.}(2014)\citenamefont{Koenig, Doganov, Schmidt, Castro~Neto, and Özyilmaz}}]{koenig_electric_2014}
\bibinfo{author}{\bibfnamefont{S.~P.} \bibnamefont{Koenig}}, \bibinfo{author}{\bibfnamefont{R.~A.} \bibnamefont{Doganov}}, \bibinfo{author}{\bibfnamefont{H.}~\bibnamefont{Schmidt}}, \bibinfo{author}{\bibfnamefont{A.~H.} \bibnamefont{Castro~Neto}}, \bibnamefont{and} \bibinfo{author}{\bibfnamefont{B.}~\bibnamefont{Özyilmaz}}, \bibinfo{journal}{Applied Physics Letters} \textbf{\bibinfo{volume}{104}} (\bibinfo{year}{2014}), ISSN \bibinfo{issn}{0003-6951, 1077-3118}, \bibinfo{note}{publisher: AIP Publishing}, \urlprefix\url{https://pubs.aip.org/apl/article/104/10/103106/1022650/Electric-field-effect-in-ultrathin-black}.

\bibitem[{\citenamefont{de~Sousa et~al.}(2024{\natexlab{b}})\citenamefont{de~Sousa, Lee, Lu, Moore, Brahlek, Wang, Bian, and Low}}]{PhysRevLett.133.146605}
\bibinfo{author}{\bibfnamefont{D.~J.~P.} \bibnamefont{de~Sousa}}, \bibinfo{author}{\bibfnamefont{S.}~\bibnamefont{Lee}}, \bibinfo{author}{\bibfnamefont{Q.}~\bibnamefont{Lu}}, \bibinfo{author}{\bibfnamefont{R.~G.} \bibnamefont{Moore}}, \bibinfo{author}{\bibfnamefont{M.}~\bibnamefont{Brahlek}}, \bibinfo{author}{\bibfnamefont{J.-P.} \bibnamefont{Wang}}, \bibinfo{author}{\bibfnamefont{G.}~\bibnamefont{Bian}}, \bibnamefont{and} \bibinfo{author}{\bibfnamefont{T.}~\bibnamefont{Low}}, \bibinfo{journal}{Phys. Rev. Lett.} \textbf{\bibinfo{volume}{133}}, \bibinfo{pages}{146605} (\bibinfo{year}{2024}{\natexlab{b}}), \urlprefix\url{https://link.aps.org/doi/10.1103/PhysRevLett.133.146605}.

\bibitem[{\citenamefont{Kowalczyk et~al.}(2020)\citenamefont{Kowalczyk, Brown, Maerkl, Lu, Chiu, Liu, Yang, Wang, Zasada, Genuzio et~al.}}]{Kowalczyk2020}
\bibinfo{author}{\bibfnamefont{P.~J.} \bibnamefont{Kowalczyk}}, \bibinfo{author}{\bibfnamefont{S.~A.} \bibnamefont{Brown}}, \bibinfo{author}{\bibfnamefont{T.}~\bibnamefont{Maerkl}}, \bibinfo{author}{\bibfnamefont{Q.}~\bibnamefont{Lu}}, \bibinfo{author}{\bibfnamefont{C.-K.} \bibnamefont{Chiu}}, \bibinfo{author}{\bibfnamefont{Y.}~\bibnamefont{Liu}}, \bibinfo{author}{\bibfnamefont{S.~A.} \bibnamefont{Yang}}, \bibinfo{author}{\bibfnamefont{X.}~\bibnamefont{Wang}}, \bibinfo{author}{\bibfnamefont{I.}~\bibnamefont{Zasada}}, \bibinfo{author}{\bibfnamefont{F.}~\bibnamefont{Genuzio}}, \bibnamefont{et~al.}, \bibinfo{journal}{ACS Nano} \textbf{\bibinfo{volume}{14}}, \bibinfo{pages}{1888–1894} (\bibinfo{year}{2020}), ISSN \bibinfo{issn}{1936-086X}, \urlprefix\url{http://dx.doi.org/10.1021/acsnano.9b08136}.

\bibitem[{\citenamefont{Du et~al.}(2022)\citenamefont{Du, Wang, Cheng, Wang, Li, Feng, Song, Shi, and He}}]{Du2022}
\bibinfo{author}{\bibfnamefont{R.}~\bibnamefont{Du}}, \bibinfo{author}{\bibfnamefont{Y.}~\bibnamefont{Wang}}, \bibinfo{author}{\bibfnamefont{M.}~\bibnamefont{Cheng}}, \bibinfo{author}{\bibfnamefont{P.}~\bibnamefont{Wang}}, \bibinfo{author}{\bibfnamefont{H.}~\bibnamefont{Li}}, \bibinfo{author}{\bibfnamefont{W.}~\bibnamefont{Feng}}, \bibinfo{author}{\bibfnamefont{L.}~\bibnamefont{Song}}, \bibinfo{author}{\bibfnamefont{J.}~\bibnamefont{Shi}}, \bibnamefont{and} \bibinfo{author}{\bibfnamefont{J.}~\bibnamefont{He}}, \bibinfo{journal}{Nature Communications} \textbf{\bibinfo{volume}{13}} (\bibinfo{year}{2022}), ISSN \bibinfo{issn}{2041-1723}, \urlprefix\url{http://dx.doi.org/10.1038/s41467-022-33917-2}.

\bibitem[{\citenamefont{Villanova et~al.}(2020)\citenamefont{Villanova, Kumar, and Barraza-Lopez}}]{PhysRevB.101.184101}
\bibinfo{author}{\bibfnamefont{J.~W.} \bibnamefont{Villanova}}, \bibinfo{author}{\bibfnamefont{P.}~\bibnamefont{Kumar}}, \bibnamefont{and} \bibinfo{author}{\bibfnamefont{S.}~\bibnamefont{Barraza-Lopez}}, \bibinfo{journal}{Phys. Rev. B} \textbf{\bibinfo{volume}{101}}, \bibinfo{pages}{184101} (\bibinfo{year}{2020}), \urlprefix\url{https://link.aps.org/doi/10.1103/PhysRevB.101.184101}.

\bibitem[{\citenamefont{Wang and Qian}(2017)}]{Wang2017}
\bibinfo{author}{\bibfnamefont{H.}~\bibnamefont{Wang}} \bibnamefont{and} \bibinfo{author}{\bibfnamefont{X.}~\bibnamefont{Qian}}, \bibinfo{journal}{2D Materials} \textbf{\bibinfo{volume}{4}}, \bibinfo{pages}{015042} (\bibinfo{year}{2017}), ISSN \bibinfo{issn}{2053-1583}, \urlprefix\url{http://dx.doi.org/10.1088/2053-1583/4/1/015042}.

\bibitem[{\citenamefont{Bao et~al.}(2019)\citenamefont{Bao, Song, Liu, Chen, Zhu, Abdelwahab, Su, Fu, Chi, Yu et~al.}}]{Bao2019}
\bibinfo{author}{\bibfnamefont{Y.}~\bibnamefont{Bao}}, \bibinfo{author}{\bibfnamefont{P.}~\bibnamefont{Song}}, \bibinfo{author}{\bibfnamefont{Y.}~\bibnamefont{Liu}}, \bibinfo{author}{\bibfnamefont{Z.}~\bibnamefont{Chen}}, \bibinfo{author}{\bibfnamefont{M.}~\bibnamefont{Zhu}}, \bibinfo{author}{\bibfnamefont{I.}~\bibnamefont{Abdelwahab}}, \bibinfo{author}{\bibfnamefont{J.}~\bibnamefont{Su}}, \bibinfo{author}{\bibfnamefont{W.}~\bibnamefont{Fu}}, \bibinfo{author}{\bibfnamefont{X.}~\bibnamefont{Chi}}, \bibinfo{author}{\bibfnamefont{W.}~\bibnamefont{Yu}}, \bibnamefont{et~al.}, \bibinfo{journal}{Nano Letters} \textbf{\bibinfo{volume}{19}}, \bibinfo{pages}{5109–5117} (\bibinfo{year}{2019}), ISSN \bibinfo{issn}{1530-6992}, \urlprefix\url{http://dx.doi.org/10.1021/acs.nanolett.9b01419}.

\bibitem[{\citenamefont{Srivastava et~al.}(2021)\citenamefont{Srivastava, Hassan, de~Sousa, Gebredingle, Joe, Ali, Zheng, Yoo, Ghosh, Teherani et~al.}}]{Srivastava2021}
\bibinfo{author}{\bibfnamefont{P.~K.} \bibnamefont{Srivastava}}, \bibinfo{author}{\bibfnamefont{Y.}~\bibnamefont{Hassan}}, \bibinfo{author}{\bibfnamefont{D.~J.~P.} \bibnamefont{de~Sousa}}, \bibinfo{author}{\bibfnamefont{Y.}~\bibnamefont{Gebredingle}}, \bibinfo{author}{\bibfnamefont{M.}~\bibnamefont{Joe}}, \bibinfo{author}{\bibfnamefont{F.}~\bibnamefont{Ali}}, \bibinfo{author}{\bibfnamefont{Y.}~\bibnamefont{Zheng}}, \bibinfo{author}{\bibfnamefont{W.~J.} \bibnamefont{Yoo}}, \bibinfo{author}{\bibfnamefont{S.}~\bibnamefont{Ghosh}}, \bibinfo{author}{\bibfnamefont{J.~T.} \bibnamefont{Teherani}}, \bibnamefont{et~al.}, \bibinfo{journal}{Nature Electronics} \textbf{\bibinfo{volume}{4}}, \bibinfo{pages}{269–276} (\bibinfo{year}{2021}), ISSN \bibinfo{issn}{2520-1131}, \urlprefix\url{http://dx.doi.org/10.1038/s41928-021-00549-1}.

\bibitem[{\citenamefont{Huang et~al.}(2024)\citenamefont{Huang, Yu, Ma, Pan, Ma, Zhou, Ma, Yang, Wu, Lei et~al.}}]{Huang2024}
\bibinfo{author}{\bibfnamefont{S.}~\bibnamefont{Huang}}, \bibinfo{author}{\bibfnamefont{B.}~\bibnamefont{Yu}}, \bibinfo{author}{\bibfnamefont{Y.}~\bibnamefont{Ma}}, \bibinfo{author}{\bibfnamefont{C.}~\bibnamefont{Pan}}, \bibinfo{author}{\bibfnamefont{J.}~\bibnamefont{Ma}}, \bibinfo{author}{\bibfnamefont{Y.}~\bibnamefont{Zhou}}, \bibinfo{author}{\bibfnamefont{Y.}~\bibnamefont{Ma}}, \bibinfo{author}{\bibfnamefont{K.}~\bibnamefont{Yang}}, \bibinfo{author}{\bibfnamefont{H.}~\bibnamefont{Wu}}, \bibinfo{author}{\bibfnamefont{Y.}~\bibnamefont{Lei}}, \bibnamefont{et~al.}, \bibinfo{journal}{Science} \textbf{\bibinfo{volume}{386}}, \bibinfo{pages}{526–531} (\bibinfo{year}{2024}), ISSN \bibinfo{issn}{1095-9203}, \urlprefix\url{http://dx.doi.org/10.1126/science.adq2977}.

\bibitem[{\citenamefont{Soltero et~al.}(2022)\citenamefont{Soltero, Guerrero-S\'anchez, Mireles, and Ruiz-Tijerina}}]{PhysRevB.105.235421}
\bibinfo{author}{\bibfnamefont{I.}~\bibnamefont{Soltero}}, \bibinfo{author}{\bibfnamefont{J.}~\bibnamefont{Guerrero-S\'anchez}}, \bibinfo{author}{\bibfnamefont{F.}~\bibnamefont{Mireles}}, \bibnamefont{and} \bibinfo{author}{\bibfnamefont{D.~A.} \bibnamefont{Ruiz-Tijerina}}, \bibinfo{journal}{Phys. Rev. B} \textbf{\bibinfo{volume}{105}}, \bibinfo{pages}{235421} (\bibinfo{year}{2022}), \urlprefix\url{https://link.aps.org/doi/10.1103/PhysRevB.105.235421}.

\bibitem[{\citenamefont{Chen et~al.}(2024)\citenamefont{Chen, Liang, Miao, Yu, Wang, Zhang, Wang, Wang, Cheng, Long et~al.}}]{Chen2024}
\bibinfo{author}{\bibfnamefont{S.}~\bibnamefont{Chen}}, \bibinfo{author}{\bibfnamefont{Z.}~\bibnamefont{Liang}}, \bibinfo{author}{\bibfnamefont{J.}~\bibnamefont{Miao}}, \bibinfo{author}{\bibfnamefont{X.-L.} \bibnamefont{Yu}}, \bibinfo{author}{\bibfnamefont{S.}~\bibnamefont{Wang}}, \bibinfo{author}{\bibfnamefont{Y.}~\bibnamefont{Zhang}}, \bibinfo{author}{\bibfnamefont{H.}~\bibnamefont{Wang}}, \bibinfo{author}{\bibfnamefont{Y.}~\bibnamefont{Wang}}, \bibinfo{author}{\bibfnamefont{C.}~\bibnamefont{Cheng}}, \bibinfo{author}{\bibfnamefont{G.}~\bibnamefont{Long}}, \bibnamefont{et~al.}, \bibinfo{journal}{Nature Communications} \textbf{\bibinfo{volume}{15}} (\bibinfo{year}{2024}), ISSN \bibinfo{issn}{2041-1723}, \urlprefix\url{http://dx.doi.org/10.1038/s41467-024-53125-4}.

\bibitem[{\citenamefont{Zhu and Long}(2023)}]{Zhu2023}
\bibinfo{author}{\bibfnamefont{Y.}~\bibnamefont{Zhu}} \bibnamefont{and} \bibinfo{author}{\bibfnamefont{R.}~\bibnamefont{Long}}, \bibinfo{journal}{Journal of Materials Chemistry A} \textbf{\bibinfo{volume}{11}}, \bibinfo{pages}{14005–14014} (\bibinfo{year}{2023}), ISSN \bibinfo{issn}{2050-7496}, \urlprefix\url{http://dx.doi.org/10.1039/D3TA02198J}.

\bibitem[{\citenamefont{Xiong et~al.}(2022)\citenamefont{Xiong, Wang, Zhu, Xu, Wu, Chen, Ma, Liu, Chen, Watanabe et~al.}}]{Xiong2022}
\bibinfo{author}{\bibfnamefont{Y.}~\bibnamefont{Xiong}}, \bibinfo{author}{\bibfnamefont{Y.}~\bibnamefont{Wang}}, \bibinfo{author}{\bibfnamefont{R.}~\bibnamefont{Zhu}}, \bibinfo{author}{\bibfnamefont{H.}~\bibnamefont{Xu}}, \bibinfo{author}{\bibfnamefont{C.}~\bibnamefont{Wu}}, \bibinfo{author}{\bibfnamefont{J.}~\bibnamefont{Chen}}, \bibinfo{author}{\bibfnamefont{Y.}~\bibnamefont{Ma}}, \bibinfo{author}{\bibfnamefont{Y.}~\bibnamefont{Liu}}, \bibinfo{author}{\bibfnamefont{Y.}~\bibnamefont{Chen}}, \bibinfo{author}{\bibfnamefont{K.}~\bibnamefont{Watanabe}}, \bibnamefont{et~al.}, \bibinfo{journal}{Science Advances} \textbf{\bibinfo{volume}{8}} (\bibinfo{year}{2022}), ISSN \bibinfo{issn}{2375-2548}, \urlprefix\url{http://dx.doi.org/10.1126/sciadv.abo0375}.

\bibitem[{\citenamefont{Wang et~al.}(2021)\citenamefont{Wang, Huang, Guo, Guo, Tian, and Liu}}]{Wang2021}
\bibinfo{author}{\bibfnamefont{S.-Y.} \bibnamefont{Wang}}, \bibinfo{author}{\bibfnamefont{K.-X.} \bibnamefont{Huang}}, \bibinfo{author}{\bibfnamefont{Q.-Q.} \bibnamefont{Guo}}, \bibinfo{author}{\bibfnamefont{H.-W.} \bibnamefont{Guo}}, \bibinfo{author}{\bibfnamefont{J.-G.} \bibnamefont{Tian}}, \bibnamefont{and} \bibinfo{author}{\bibfnamefont{Z.-B.} \bibnamefont{Liu}}, \bibinfo{journal}{The Journal of Physical Chemistry Letters} \textbf{\bibinfo{volume}{12}}, \bibinfo{pages}{4755–4761} (\bibinfo{year}{2021}), ISSN \bibinfo{issn}{1948-7185}, \urlprefix\url{http://dx.doi.org/10.1021/acs.jpclett.1c01029}.

\bibitem[{\citenamefont{Wang and Zou}(2022)}]{2Wang2022}
\bibinfo{author}{\bibfnamefont{E.}~\bibnamefont{Wang}} \bibnamefont{and} \bibinfo{author}{\bibfnamefont{X.}~\bibnamefont{Zou}}, \bibinfo{journal}{Nanoscale} \textbf{\bibinfo{volume}{14}}, \bibinfo{pages}{3758–3767} (\bibinfo{year}{2022}), ISSN \bibinfo{issn}{2040-3372}, \urlprefix\url{http://dx.doi.org/10.1039/d1nr07736h}.

\bibitem[{\citenamefont{He and Weng}(2021)}]{He2021}
\bibinfo{author}{\bibfnamefont{Z.}~\bibnamefont{He}} \bibnamefont{and} \bibinfo{author}{\bibfnamefont{H.}~\bibnamefont{Weng}}, \bibinfo{journal}{npj Quantum Materials} \textbf{\bibinfo{volume}{6}} (\bibinfo{year}{2021}), ISSN \bibinfo{issn}{2397-4648}, \urlprefix\url{http://dx.doi.org/10.1038/s41535-021-00403-9}.

\bibitem[{\citenamefont{Sevik et~al.}(2017)\citenamefont{Sevik, Wallbank, G\"{u}lseren, Peeters, and undefinedakır}}]{Sevik2017}
\bibinfo{author}{\bibfnamefont{C.}~\bibnamefont{Sevik}}, \bibinfo{author}{\bibfnamefont{J.~R.} \bibnamefont{Wallbank}}, \bibinfo{author}{\bibfnamefont{O.}~\bibnamefont{G\"{u}lseren}}, \bibinfo{author}{\bibfnamefont{F.~M.} \bibnamefont{Peeters}}, \bibnamefont{and} \bibinfo{author}{\bibfnamefont{D.}~\bibnamefont{undefinedakır}}, \bibinfo{journal}{2D Materials} \textbf{\bibinfo{volume}{4}}, \bibinfo{pages}{035025} (\bibinfo{year}{2017}), ISSN \bibinfo{issn}{2053-1583}, \urlprefix\url{http://dx.doi.org/10.1088/2053-1583/aa80c4}.

\bibitem[{\citenamefont{Guo et~al.}(2023)\citenamefont{Guo, Zhang, and Lu}}]{Guo2023}
\bibinfo{author}{\bibfnamefont{H.}~\bibnamefont{Guo}}, \bibinfo{author}{\bibfnamefont{X.}~\bibnamefont{Zhang}}, \bibnamefont{and} \bibinfo{author}{\bibfnamefont{G.}~\bibnamefont{Lu}}, \bibinfo{journal}{Science Advances} \textbf{\bibinfo{volume}{9}} (\bibinfo{year}{2023}), ISSN \bibinfo{issn}{2375-2548}, \urlprefix\url{http://dx.doi.org/10.1126/sciadv.adi5404}.

\bibitem[{\citenamefont{Kang et~al.}(2017)\citenamefont{Kang, Zhang, Michaud-Rioux, Kong, Hu, Yu, and Guo}}]{PhysRevB.96.195406}
\bibinfo{author}{\bibfnamefont{P.}~\bibnamefont{Kang}}, \bibinfo{author}{\bibfnamefont{W.-T.} \bibnamefont{Zhang}}, \bibinfo{author}{\bibfnamefont{V.}~\bibnamefont{Michaud-Rioux}}, \bibinfo{author}{\bibfnamefont{X.-H.} \bibnamefont{Kong}}, \bibinfo{author}{\bibfnamefont{C.}~\bibnamefont{Hu}}, \bibinfo{author}{\bibfnamefont{G.-H.} \bibnamefont{Yu}}, \bibnamefont{and} \bibinfo{author}{\bibfnamefont{H.}~\bibnamefont{Guo}}, \bibinfo{journal}{Phys. Rev. B} \textbf{\bibinfo{volume}{96}}, \bibinfo{pages}{195406} (\bibinfo{year}{2017}), \urlprefix\url{https://link.aps.org/doi/10.1103/PhysRevB.96.195406}.

\bibitem[{\citenamefont{Cao et~al.}(2016)\citenamefont{Cao, Li, Qiu, and Louie}}]{Cao2016}
\bibinfo{author}{\bibfnamefont{T.}~\bibnamefont{Cao}}, \bibinfo{author}{\bibfnamefont{Z.}~\bibnamefont{Li}}, \bibinfo{author}{\bibfnamefont{D.~Y.} \bibnamefont{Qiu}}, \bibnamefont{and} \bibinfo{author}{\bibfnamefont{S.~G.} \bibnamefont{Louie}}, \bibinfo{journal}{Nano Letters} \textbf{\bibinfo{volume}{16}}, \bibinfo{pages}{5542–5546} (\bibinfo{year}{2016}), ISSN \bibinfo{issn}{1530-6992}, \urlprefix\url{http://dx.doi.org/10.1021/acs.nanolett.6b02084}.

\bibitem[{\citenamefont{Kim et~al.}(2016)\citenamefont{Kim, Sánchez-Castillo, Ziegler, Ogawa, Noguez, and Park}}]{Kim2016}
\bibinfo{author}{\bibfnamefont{C.-J.} \bibnamefont{Kim}}, \bibinfo{author}{\bibfnamefont{A.}~\bibnamefont{Sánchez-Castillo}}, \bibinfo{author}{\bibfnamefont{Z.}~\bibnamefont{Ziegler}}, \bibinfo{author}{\bibfnamefont{Y.}~\bibnamefont{Ogawa}}, \bibinfo{author}{\bibfnamefont{C.}~\bibnamefont{Noguez}}, \bibnamefont{and} \bibinfo{author}{\bibfnamefont{J.}~\bibnamefont{Park}}, \bibinfo{journal}{Nature Nanotechnology} \textbf{\bibinfo{volume}{11}}, \bibinfo{pages}{520–524} (\bibinfo{year}{2016}), ISSN \bibinfo{issn}{1748-3395}, \urlprefix\url{http://dx.doi.org/10.1038/nnano.2016.3}.

\bibitem[{\citenamefont{Low et~al.}(2015)\citenamefont{Low, Jiang, and Guinea}}]{low_topological_2015}
\bibinfo{author}{\bibfnamefont{T.}~\bibnamefont{Low}}, \bibinfo{author}{\bibfnamefont{Y.}~\bibnamefont{Jiang}}, \bibnamefont{and} \bibinfo{author}{\bibfnamefont{F.}~\bibnamefont{Guinea}}, \bibinfo{journal}{Physical Review B} \textbf{\bibinfo{volume}{92}} (\bibinfo{year}{2015}), ISSN \bibinfo{issn}{1098-0121, 1550-235X}, \bibinfo{note}{publisher: American Physical Society (APS)}, \urlprefix\url{https://link.aps.org/doi/10.1103/PhysRevB.92.235447}.

\bibitem[{\citenamefont{Forte et~al.}(2019)\citenamefont{Forte, de~Sousa, and Pereira}}]{Forte2019}
\bibinfo{author}{\bibfnamefont{J.~D.} \bibnamefont{Forte}}, \bibinfo{author}{\bibfnamefont{D.~J.} \bibnamefont{de~Sousa}}, \bibnamefont{and} \bibinfo{author}{\bibfnamefont{J.~M.} \bibnamefont{Pereira}}, \bibinfo{journal}{Physica E: Low-dimensional Systems and Nanostructures} \textbf{\bibinfo{volume}{114}}, \bibinfo{pages}{113578} (\bibinfo{year}{2019}), ISSN \bibinfo{issn}{1386-9477}, \urlprefix\url{http://dx.doi.org/10.1016/j.physe.2019.113578}.

\bibitem[{\citenamefont{Li et~al.}(2018)\citenamefont{Li, Partoens, and Peeters}}]{PhysRevB.97.155424}
\bibinfo{author}{\bibfnamefont{L.~L.} \bibnamefont{Li}}, \bibinfo{author}{\bibfnamefont{B.}~\bibnamefont{Partoens}}, \bibnamefont{and} \bibinfo{author}{\bibfnamefont{F.~M.} \bibnamefont{Peeters}}, \bibinfo{journal}{Phys. Rev. B} \textbf{\bibinfo{volume}{97}}, \bibinfo{pages}{155424} (\bibinfo{year}{2018}), \urlprefix\url{https://link.aps.org/doi/10.1103/PhysRevB.97.155424}.

\bibitem[{\citenamefont{Van~Veen et~al.}(2019)\citenamefont{Van~Veen, Nemilentsau, Kumar, Roldán, Katsnelson, Low, and Yuan}}]{van_veen_tuning_2019}
\bibinfo{author}{\bibfnamefont{E.}~\bibnamefont{Van~Veen}}, \bibinfo{author}{\bibfnamefont{A.}~\bibnamefont{Nemilentsau}}, \bibinfo{author}{\bibfnamefont{A.}~\bibnamefont{Kumar}}, \bibinfo{author}{\bibfnamefont{R.}~\bibnamefont{Roldán}}, \bibinfo{author}{\bibfnamefont{M.~I.} \bibnamefont{Katsnelson}}, \bibinfo{author}{\bibfnamefont{T.}~\bibnamefont{Low}}, \bibnamefont{and} \bibinfo{author}{\bibfnamefont{S.}~\bibnamefont{Yuan}}, \bibinfo{journal}{Physical Review Applied} \textbf{\bibinfo{volume}{12}} (\bibinfo{year}{2019}), ISSN \bibinfo{issn}{2331-7019}, \bibinfo{note}{publisher: American Physical Society (APS)}, \urlprefix\url{https://link.aps.org/doi/10.1103/PhysRevApplied.12.014011}.

\bibitem[{\citenamefont{Low et~al.}(2014{\natexlab{a}})\citenamefont{Low, Rodin, Carvalho, Jiang, Wang, Xia, and Castro~Neto}}]{low_tunable_2014}
\bibinfo{author}{\bibfnamefont{T.}~\bibnamefont{Low}}, \bibinfo{author}{\bibfnamefont{A.~S.} \bibnamefont{Rodin}}, \bibinfo{author}{\bibfnamefont{A.}~\bibnamefont{Carvalho}}, \bibinfo{author}{\bibfnamefont{Y.}~\bibnamefont{Jiang}}, \bibinfo{author}{\bibfnamefont{H.}~\bibnamefont{Wang}}, \bibinfo{author}{\bibfnamefont{F.}~\bibnamefont{Xia}}, \bibnamefont{and} \bibinfo{author}{\bibfnamefont{A.~H.} \bibnamefont{Castro~Neto}}, \bibinfo{journal}{Physical Review B} \textbf{\bibinfo{volume}{90}} (\bibinfo{year}{2014}{\natexlab{a}}), ISSN \bibinfo{issn}{1098-0121, 1550-235X}, \bibinfo{note}{publisher: American Physical Society (APS)}, \urlprefix\url{https://link.aps.org/doi/10.1103/PhysRevB.90.075434}.

\bibitem[{\citenamefont{Chaves et~al.}(2015)\citenamefont{Chaves, Low, Avouris, Çakır, and Peeters}}]{chaves_anisotropic_2015}
\bibinfo{author}{\bibfnamefont{A.}~\bibnamefont{Chaves}}, \bibinfo{author}{\bibfnamefont{T.}~\bibnamefont{Low}}, \bibinfo{author}{\bibfnamefont{P.}~\bibnamefont{Avouris}}, \bibinfo{author}{\bibfnamefont{D.}~\bibnamefont{Çakır}}, \bibnamefont{and} \bibinfo{author}{\bibfnamefont{F.~M.} \bibnamefont{Peeters}}, \bibinfo{journal}{Physical Review B} \textbf{\bibinfo{volume}{91}} (\bibinfo{year}{2015}), ISSN \bibinfo{issn}{1098-0121, 1550-235X}, \bibinfo{note}{publisher: American Physical Society (APS)}, \urlprefix\url{https://link.aps.org/doi/10.1103/PhysRevB.91.155311}.

\bibitem[{\citenamefont{Zhang et~al.}(2017)\citenamefont{Zhang, Huang, Chaves, Song, Özçelik, Low, and Yan}}]{zhang_infrared_2017}
\bibinfo{author}{\bibfnamefont{G.}~\bibnamefont{Zhang}}, \bibinfo{author}{\bibfnamefont{S.}~\bibnamefont{Huang}}, \bibinfo{author}{\bibfnamefont{A.}~\bibnamefont{Chaves}}, \bibinfo{author}{\bibfnamefont{C.}~\bibnamefont{Song}}, \bibinfo{author}{\bibfnamefont{V.~O.} \bibnamefont{Özçelik}}, \bibinfo{author}{\bibfnamefont{T.}~\bibnamefont{Low}}, \bibnamefont{and} \bibinfo{author}{\bibfnamefont{H.}~\bibnamefont{Yan}}, \bibinfo{journal}{Nature Communications} \textbf{\bibinfo{volume}{8}} (\bibinfo{year}{2017}), ISSN \bibinfo{issn}{2041-1723}, \bibinfo{note}{publisher: Springer Science and Business Media LLC}, \urlprefix\url{https://www.nature.com/articles/ncomms14071}.

\bibitem[{\citenamefont{Arra et~al.}(2019)\citenamefont{Arra, Babar, and Kabir}}]{arra_exciton_2019}
\bibinfo{author}{\bibfnamefont{S.}~\bibnamefont{Arra}}, \bibinfo{author}{\bibfnamefont{R.}~\bibnamefont{Babar}}, \bibnamefont{and} \bibinfo{author}{\bibfnamefont{M.}~\bibnamefont{Kabir}}, \bibinfo{journal}{Physical Review B} \textbf{\bibinfo{volume}{99}} (\bibinfo{year}{2019}), ISSN \bibinfo{issn}{2469-9950, 2469-9969}, \bibinfo{note}{publisher: American Physical Society (APS)}, \urlprefix\url{https://link.aps.org/doi/10.1103/PhysRevB.99.045432}.

\bibitem[{\citenamefont{Zhang et~al.}(2018)\citenamefont{Zhang, Chaves, Huang, Wang, Xing, Low, and Yan}}]{zhang_determination_2018}
\bibinfo{author}{\bibfnamefont{G.}~\bibnamefont{Zhang}}, \bibinfo{author}{\bibfnamefont{A.}~\bibnamefont{Chaves}}, \bibinfo{author}{\bibfnamefont{S.}~\bibnamefont{Huang}}, \bibinfo{author}{\bibfnamefont{F.}~\bibnamefont{Wang}}, \bibinfo{author}{\bibfnamefont{Q.}~\bibnamefont{Xing}}, \bibinfo{author}{\bibfnamefont{T.}~\bibnamefont{Low}}, \bibnamefont{and} \bibinfo{author}{\bibfnamefont{H.}~\bibnamefont{Yan}}, \bibinfo{journal}{Science Advances} \textbf{\bibinfo{volume}{4}} (\bibinfo{year}{2018}), ISSN \bibinfo{issn}{2375-2548}, \bibinfo{note}{publisher: American Association for the Advancement of Science (AAAS)}, \urlprefix\url{https://www.science.org/doi/10.1126/sciadv.aap9977}.

\bibitem[{\citenamefont{Qiu et~al.}(2017)\citenamefont{Qiu, Da~Jornada, and Louie}}]{qiu_environmental_2017}
\bibinfo{author}{\bibfnamefont{D.~Y.} \bibnamefont{Qiu}}, \bibinfo{author}{\bibfnamefont{F.~H.} \bibnamefont{Da~Jornada}}, \bibnamefont{and} \bibinfo{author}{\bibfnamefont{S.~G.} \bibnamefont{Louie}}, \bibinfo{journal}{Nano Letters} \textbf{\bibinfo{volume}{17}}, \bibinfo{pages}{4706} (\bibinfo{year}{2017}), ISSN \bibinfo{issn}{1530-6984, 1530-6992}, \bibinfo{note}{publisher: American Chemical Society (ACS)}, \urlprefix\url{https://pubs.acs.org/doi/10.1021/acs.nanolett.7b01365}.

\bibitem[{\citenamefont{Ghosh et~al.}(2017)\citenamefont{Ghosh, Kumar, Thakur, Chauhan, Bhowmick, and Agarwal}}]{ghosh_anisotropic_2017}
\bibinfo{author}{\bibfnamefont{B.}~\bibnamefont{Ghosh}}, \bibinfo{author}{\bibfnamefont{P.}~\bibnamefont{Kumar}}, \bibinfo{author}{\bibfnamefont{A.}~\bibnamefont{Thakur}}, \bibinfo{author}{\bibfnamefont{Y.~S.} \bibnamefont{Chauhan}}, \bibinfo{author}{\bibfnamefont{S.}~\bibnamefont{Bhowmick}}, \bibnamefont{and} \bibinfo{author}{\bibfnamefont{A.}~\bibnamefont{Agarwal}}, \bibinfo{journal}{Physical Review B} \textbf{\bibinfo{volume}{96}} (\bibinfo{year}{2017}), ISSN \bibinfo{issn}{2469-9950, 2469-9969}, \bibinfo{note}{publisher: American Physical Society (APS)}, \urlprefix\url{http://link.aps.org/doi/10.1103/PhysRevB.96.035422}.

\bibitem[{\citenamefont{Villegas et~al.}(2016)\citenamefont{Villegas, Rodin, Carvalho, and Rocha}}]{villegas_two-dimensional_2016}
\bibinfo{author}{\bibfnamefont{C.~E.~P.} \bibnamefont{Villegas}}, \bibinfo{author}{\bibfnamefont{A.~S.} \bibnamefont{Rodin}}, \bibinfo{author}{\bibfnamefont{A.}~\bibnamefont{Carvalho}}, \bibnamefont{and} \bibinfo{author}{\bibfnamefont{A.~R.} \bibnamefont{Rocha}}, \bibinfo{journal}{Physical Chemistry Chemical Physics} \textbf{\bibinfo{volume}{18}}, \bibinfo{pages}{27829} (\bibinfo{year}{2016}), ISSN \bibinfo{issn}{1463-9076, 1463-9084}, \bibinfo{note}{publisher: Royal Society of Chemistry (RSC)}, \urlprefix\url{https://xlink.rsc.org/?DOI=C6CP05566D}.

\bibitem[{\citenamefont{Correas-Serrano et~al.}(2016)\citenamefont{Correas-Serrano, Gomez-Diaz, Melcon, and Alù}}]{correas-serrano_black_2016}
\bibinfo{author}{\bibfnamefont{D.}~\bibnamefont{Correas-Serrano}}, \bibinfo{author}{\bibfnamefont{J.~S.} \bibnamefont{Gomez-Diaz}}, \bibinfo{author}{\bibfnamefont{A.~A.} \bibnamefont{Melcon}}, \bibnamefont{and} \bibinfo{author}{\bibfnamefont{A.}~\bibnamefont{Alù}}, \bibinfo{journal}{Journal of Optics} \textbf{\bibinfo{volume}{18}}, \bibinfo{pages}{104006} (\bibinfo{year}{2016}), ISSN \bibinfo{issn}{2040-8978, 2040-8986}, \bibinfo{note}{publisher: IOP Publishing}, \urlprefix\url{https://iopscience.iop.org/article/10.1088/2040-8978/18/10/104006}.

\bibitem[{\citenamefont{Surrente et~al.}(2016)\citenamefont{Surrente, Mitioglu, Galkowski, Tabis, Maude, and Plochocka}}]{surrente_excitons_2016}
\bibinfo{author}{\bibfnamefont{A.}~\bibnamefont{Surrente}}, \bibinfo{author}{\bibfnamefont{A.~A.} \bibnamefont{Mitioglu}}, \bibinfo{author}{\bibfnamefont{K.}~\bibnamefont{Galkowski}}, \bibinfo{author}{\bibfnamefont{W.}~\bibnamefont{Tabis}}, \bibinfo{author}{\bibfnamefont{D.~K.} \bibnamefont{Maude}}, \bibnamefont{and} \bibinfo{author}{\bibfnamefont{P.}~\bibnamefont{Plochocka}}, \bibinfo{journal}{Physical Review B} \textbf{\bibinfo{volume}{93}} (\bibinfo{year}{2016}), ISSN \bibinfo{issn}{2469-9950, 2469-9969}, \bibinfo{note}{publisher: American Physical Society (APS)}, \urlprefix\url{https://link.aps.org/doi/10.1103/PhysRevB.93.121405}.

\bibitem[{\citenamefont{Nemilentsau et~al.}(2016)\citenamefont{Nemilentsau, Low, and Hanson}}]{nemilentsau_anisotropic_2016}
\bibinfo{author}{\bibfnamefont{A.}~\bibnamefont{Nemilentsau}}, \bibinfo{author}{\bibfnamefont{T.}~\bibnamefont{Low}}, \bibnamefont{and} \bibinfo{author}{\bibfnamefont{G.}~\bibnamefont{Hanson}}, \bibinfo{journal}{Physical Review Letters} \textbf{\bibinfo{volume}{116}} (\bibinfo{year}{2016}), ISSN \bibinfo{issn}{0031-9007, 1079-7114}, \bibinfo{note}{publisher: American Physical Society (APS)}, \urlprefix\url{https://link.aps.org/doi/10.1103/PhysRevLett.116.066804}.

\bibitem[{\citenamefont{Tran et~al.}(2015)\citenamefont{Tran, Fei, and Yang}}]{tran_quasiparticle_2015}
\bibinfo{author}{\bibfnamefont{V.}~\bibnamefont{Tran}}, \bibinfo{author}{\bibfnamefont{R.}~\bibnamefont{Fei}}, \bibnamefont{and} \bibinfo{author}{\bibfnamefont{L.}~\bibnamefont{Yang}}, \bibinfo{journal}{2D Materials} \textbf{\bibinfo{volume}{2}}, \bibinfo{pages}{044014} (\bibinfo{year}{2015}), ISSN \bibinfo{issn}{2053-1583}, \bibinfo{note}{publisher: IOP Publishing}, \urlprefix\url{https://iopscience.iop.org/article/10.1088/2053-1583/2/4/044014}.

\bibitem[{\citenamefont{Wang et~al.}(2015)\citenamefont{Wang, Jones, Seyler, Tran, Jia, Zhao, Wang, Yang, Xu, and Xia}}]{wang_highly_2015}
\bibinfo{author}{\bibfnamefont{X.}~\bibnamefont{Wang}}, \bibinfo{author}{\bibfnamefont{A.~M.} \bibnamefont{Jones}}, \bibinfo{author}{\bibfnamefont{K.~L.} \bibnamefont{Seyler}}, \bibinfo{author}{\bibfnamefont{V.}~\bibnamefont{Tran}}, \bibinfo{author}{\bibfnamefont{Y.}~\bibnamefont{Jia}}, \bibinfo{author}{\bibfnamefont{H.}~\bibnamefont{Zhao}}, \bibinfo{author}{\bibfnamefont{H.}~\bibnamefont{Wang}}, \bibinfo{author}{\bibfnamefont{L.}~\bibnamefont{Yang}}, \bibinfo{author}{\bibfnamefont{X.}~\bibnamefont{Xu}}, \bibnamefont{and} \bibinfo{author}{\bibfnamefont{F.}~\bibnamefont{Xia}}, \bibinfo{journal}{Nature Nanotechnology} \textbf{\bibinfo{volume}{10}}, \bibinfo{pages}{517} (\bibinfo{year}{2015}), ISSN \bibinfo{issn}{1748-3387, 1748-3395}, \bibinfo{note}{publisher: Springer Science and Business Media LLC}, \urlprefix\url{https://www.nature.com/articles/nnano.2015.71}.

\bibitem[{\citenamefont{Low et~al.}(2014{\natexlab{b}})\citenamefont{Low, Roldán, Wang, Xia, Avouris, Moreno, and Guinea}}]{low_plasmons_2014}
\bibinfo{author}{\bibfnamefont{T.}~\bibnamefont{Low}}, \bibinfo{author}{\bibfnamefont{R.}~\bibnamefont{Roldán}}, \bibinfo{author}{\bibfnamefont{H.}~\bibnamefont{Wang}}, \bibinfo{author}{\bibfnamefont{F.}~\bibnamefont{Xia}}, \bibinfo{author}{\bibfnamefont{P.}~\bibnamefont{Avouris}}, \bibinfo{author}{\bibfnamefont{L.~M.} \bibnamefont{Moreno}}, \bibnamefont{and} \bibinfo{author}{\bibfnamefont{F.}~\bibnamefont{Guinea}}, \bibinfo{journal}{Physical Review Letters} \textbf{\bibinfo{volume}{113}} (\bibinfo{year}{2014}{\natexlab{b}}), ISSN \bibinfo{issn}{0031-9007, 1079-7114}, \bibinfo{note}{publisher: American Physical Society (APS)}, \urlprefix\url{https://link.aps.org/doi/10.1103/PhysRevLett.113.106802}.

\bibitem[{\citenamefont{Sinner et~al.}(2023{\natexlab{a}})\citenamefont{Sinner, Pantale\'on, and Guinea}}]{PhysRevLett.131.166402}
\bibinfo{author}{\bibfnamefont{A.}~\bibnamefont{Sinner}}, \bibinfo{author}{\bibfnamefont{P.~A.} \bibnamefont{Pantale\'on}}, \bibnamefont{and} \bibinfo{author}{\bibfnamefont{F.}~\bibnamefont{Guinea}}, \bibinfo{journal}{Phys. Rev. Lett.} \textbf{\bibinfo{volume}{131}}, \bibinfo{pages}{166402} (\bibinfo{year}{2023}{\natexlab{a}}), \urlprefix\url{https://link.aps.org/doi/10.1103/PhysRevLett.131.166402}.

\bibitem[{Sno()}]{Snote}
\bibinfo{note}{See Supplemental Material at http://link/supplemental/XXX for details on the continuum model and first principles calculations.}

\bibitem[{\citenamefont{Tomonaga}(1950)}]{Tomonaga1950}
\bibinfo{author}{\bibfnamefont{S.-i.} \bibnamefont{Tomonaga}}, \bibinfo{journal}{Progress of Theoretical Physics} \textbf{\bibinfo{volume}{5}}, \bibinfo{pages}{544–569} (\bibinfo{year}{1950}), ISSN \bibinfo{issn}{1347-4081}, \urlprefix\url{http://dx.doi.org/10.1143/ptp/5.4.544}.

\bibitem[{\citenamefont{Luttinger}(1963)}]{Luttinger1963}
\bibinfo{author}{\bibfnamefont{J.~M.} \bibnamefont{Luttinger}}, \bibinfo{journal}{Journal of Mathematical Physics} \textbf{\bibinfo{volume}{4}}, \bibinfo{pages}{1154–1162} (\bibinfo{year}{1963}), ISSN \bibinfo{issn}{1089-7658}, \urlprefix\url{http://dx.doi.org/10.1063/1.1704046}.

\bibitem[{\citenamefont{Imambekov et~al.}(2012)\citenamefont{Imambekov, Schmidt, and Glazman}}]{RevModPhys.84.1253}
\bibinfo{author}{\bibfnamefont{A.}~\bibnamefont{Imambekov}}, \bibinfo{author}{\bibfnamefont{T.~L.} \bibnamefont{Schmidt}}, \bibnamefont{and} \bibinfo{author}{\bibfnamefont{L.~I.} \bibnamefont{Glazman}}, \bibinfo{journal}{Rev. Mod. Phys.} \textbf{\bibinfo{volume}{84}}, \bibinfo{pages}{1253} (\bibinfo{year}{2012}), \urlprefix\url{https://link.aps.org/doi/10.1103/RevModPhys.84.1253}.

\bibitem[{\citenamefont{Haldane}(1981{\natexlab{a}})}]{Haldane1981}
\bibinfo{author}{\bibfnamefont{F.~D.~M.} \bibnamefont{Haldane}}, \bibinfo{journal}{Journal of Physics C: Solid State Physics} \textbf{\bibinfo{volume}{14}}, \bibinfo{pages}{2585–2609} (\bibinfo{year}{1981}{\natexlab{a}}), ISSN \bibinfo{issn}{0022-3719}, \urlprefix\url{http://dx.doi.org/10.1088/0022-3719/14/19/010}.

\bibitem[{\citenamefont{Haldane}(1980)}]{PhysRevLett.45.1358}
\bibinfo{author}{\bibfnamefont{F.~D.~M.} \bibnamefont{Haldane}}, \bibinfo{journal}{Phys. Rev. Lett.} \textbf{\bibinfo{volume}{45}}, \bibinfo{pages}{1358} (\bibinfo{year}{1980}), \urlprefix\url{https://link.aps.org/doi/10.1103/PhysRevLett.45.1358}.

\bibitem[{\citenamefont{Haldane}(1981{\natexlab{b}})}]{2Haldane1981}
\bibinfo{author}{\bibfnamefont{F.}~\bibnamefont{Haldane}}, \bibinfo{journal}{Physics Letters A} \textbf{\bibinfo{volume}{81}}, \bibinfo{pages}{153–155} (\bibinfo{year}{1981}{\natexlab{b}}), ISSN \bibinfo{issn}{0375-9601}, \urlprefix\url{http://dx.doi.org/10.1016/0375-9601(81)90049-9}.

\bibitem[{\citenamefont{Giamarchi}(2003)}]{book}
\bibinfo{author}{\bibfnamefont{T.}~\bibnamefont{Giamarchi}}, \emph{\bibinfo{title}{Quantum Physics in One Dimension}} (\bibinfo{publisher}{Oxford University Press}, \bibinfo{address}{Oxford,}, \bibinfo{year}{2003}).

\bibitem[{\citenamefont{Kane and Fisher}(1992)}]{PhysRevLett.68.1220}
\bibinfo{author}{\bibfnamefont{C.~L.} \bibnamefont{Kane}} \bibnamefont{and} \bibinfo{author}{\bibfnamefont{M.~P.~A.} \bibnamefont{Fisher}}, \bibinfo{journal}{Phys. Rev. Lett.} \textbf{\bibinfo{volume}{68}}, \bibinfo{pages}{1220} (\bibinfo{year}{1992}), \urlprefix\url{https://link.aps.org/doi/10.1103/PhysRevLett.68.1220}.

\bibitem[{\citenamefont{Jompol et~al.}(2009)\citenamefont{Jompol, Ford, Griffiths, Farrer, Jones, Anderson, Ritchie, Silk, and Schofield}}]{Jompol2009}
\bibinfo{author}{\bibfnamefont{Y.}~\bibnamefont{Jompol}}, \bibinfo{author}{\bibfnamefont{C.~J.~B.} \bibnamefont{Ford}}, \bibinfo{author}{\bibfnamefont{J.~P.} \bibnamefont{Griffiths}}, \bibinfo{author}{\bibfnamefont{I.}~\bibnamefont{Farrer}}, \bibinfo{author}{\bibfnamefont{G.~A.~C.} \bibnamefont{Jones}}, \bibinfo{author}{\bibfnamefont{D.}~\bibnamefont{Anderson}}, \bibinfo{author}{\bibfnamefont{D.~A.} \bibnamefont{Ritchie}}, \bibinfo{author}{\bibfnamefont{T.~W.} \bibnamefont{Silk}}, \bibnamefont{and} \bibinfo{author}{\bibfnamefont{A.~J.} \bibnamefont{Schofield}}, \bibinfo{journal}{Science} \textbf{\bibinfo{volume}{325}}, \bibinfo{pages}{597–601} (\bibinfo{year}{2009}), ISSN \bibinfo{issn}{1095-9203}, \urlprefix\url{http://dx.doi.org/10.1126/science.1171769}.

\bibitem[{\citenamefont{Schmidt et~al.}(2010)\citenamefont{Schmidt, Imambekov, and Glazman}}]{PhysRevB.82.245104}
\bibinfo{author}{\bibfnamefont{T.~L.} \bibnamefont{Schmidt}}, \bibinfo{author}{\bibfnamefont{A.}~\bibnamefont{Imambekov}}, \bibnamefont{and} \bibinfo{author}{\bibfnamefont{L.~I.} \bibnamefont{Glazman}}, \bibinfo{journal}{Phys. Rev. B} \textbf{\bibinfo{volume}{82}}, \bibinfo{pages}{245104} (\bibinfo{year}{2010}), \urlprefix\url{https://link.aps.org/doi/10.1103/PhysRevB.82.245104}.

\bibitem[{\citenamefont{Vishwanath and Carpentier}(2001)}]{PhysRevLett.86.676}
\bibinfo{author}{\bibfnamefont{A.}~\bibnamefont{Vishwanath}} \bibnamefont{and} \bibinfo{author}{\bibfnamefont{D.}~\bibnamefont{Carpentier}}, \bibinfo{journal}{Phys. Rev. Lett.} \textbf{\bibinfo{volume}{86}}, \bibinfo{pages}{676} (\bibinfo{year}{2001}), \urlprefix\url{https://link.aps.org/doi/10.1103/PhysRevLett.86.676}.

\bibitem[{\citenamefont{Mukhopadhyay et~al.}(2001)\citenamefont{Mukhopadhyay, Kane, and Lubensky}}]{PhysRevB.64.045120}
\bibinfo{author}{\bibfnamefont{R.}~\bibnamefont{Mukhopadhyay}}, \bibinfo{author}{\bibfnamefont{C.~L.} \bibnamefont{Kane}}, \bibnamefont{and} \bibinfo{author}{\bibfnamefont{T.~C.} \bibnamefont{Lubensky}}, \bibinfo{journal}{Phys. Rev. B} \textbf{\bibinfo{volume}{64}}, \bibinfo{pages}{045120} (\bibinfo{year}{2001}), \urlprefix\url{https://link.aps.org/doi/10.1103/PhysRevB.64.045120}.

\bibitem[{\citenamefont{de~Sousa et~al.}(2017)\citenamefont{de~Sousa, de~Castro, da~Costa, Pereira, and Low}}]{PhysRevB.96.155427}
\bibinfo{author}{\bibfnamefont{D.~J.~P.} \bibnamefont{de~Sousa}}, \bibinfo{author}{\bibfnamefont{L.~V.} \bibnamefont{de~Castro}}, \bibinfo{author}{\bibfnamefont{D.~R.} \bibnamefont{da~Costa}}, \bibinfo{author}{\bibfnamefont{J.~M.} \bibnamefont{Pereira}}, \bibnamefont{and} \bibinfo{author}{\bibfnamefont{T.}~\bibnamefont{Low}}, \bibinfo{journal}{Phys. Rev. B} \textbf{\bibinfo{volume}{96}}, \bibinfo{pages}{155427} (\bibinfo{year}{2017}), \urlprefix\url{https://link.aps.org/doi/10.1103/PhysRevB.96.155427}.

\bibitem[{\citenamefont{Pereira and Katsnelson}(2015)}]{PhysRevB.92.075437}
\bibinfo{author}{\bibfnamefont{J.~M.} \bibnamefont{Pereira}} \bibnamefont{and} \bibinfo{author}{\bibfnamefont{M.~I.} \bibnamefont{Katsnelson}}, \bibinfo{journal}{Phys. Rev. B} \textbf{\bibinfo{volume}{92}}, \bibinfo{pages}{075437} (\bibinfo{year}{2015}), \urlprefix\url{https://link.aps.org/doi/10.1103/PhysRevB.92.075437}.

\bibitem[{\citenamefont{Bistritzer and MacDonald}(2010)}]{PhysRevB.81.245412}
\bibinfo{author}{\bibfnamefont{R.}~\bibnamefont{Bistritzer}} \bibnamefont{and} \bibinfo{author}{\bibfnamefont{A.~H.} \bibnamefont{MacDonald}}, \bibinfo{journal}{Phys. Rev. B} \textbf{\bibinfo{volume}{81}}, \bibinfo{pages}{245412} (\bibinfo{year}{2010}), \urlprefix\url{https://link.aps.org/doi/10.1103/PhysRevB.81.245412}.

\bibitem[{\citenamefont{de~Sousa et~al.}(2016)\citenamefont{de~Sousa, de~Castro, da~Costa, and Pereira}}]{PhysRevB.94.235415}
\bibinfo{author}{\bibfnamefont{D.~J.~P.} \bibnamefont{de~Sousa}}, \bibinfo{author}{\bibfnamefont{L.~V.} \bibnamefont{de~Castro}}, \bibinfo{author}{\bibfnamefont{D.~R.} \bibnamefont{da~Costa}}, \bibnamefont{and} \bibinfo{author}{\bibfnamefont{J.~M.} \bibnamefont{Pereira}}, \bibinfo{journal}{Phys. Rev. B} \textbf{\bibinfo{volume}{94}}, \bibinfo{pages}{235415} (\bibinfo{year}{2016}), \urlprefix\url{https://link.aps.org/doi/10.1103/PhysRevB.94.235415}.

\bibitem[{\citenamefont{Rudenko and Katsnelson}(2014)}]{PhysRevB.89.201408}
\bibinfo{author}{\bibfnamefont{A.~N.} \bibnamefont{Rudenko}} \bibnamefont{and} \bibinfo{author}{\bibfnamefont{M.~I.} \bibnamefont{Katsnelson}}, \bibinfo{journal}{Phys. Rev. B} \textbf{\bibinfo{volume}{89}}, \bibinfo{pages}{201408} (\bibinfo{year}{2014}), \urlprefix\url{https://link.aps.org/doi/10.1103/PhysRevB.89.201408}.

\bibitem[{\citenamefont{Rudenko et~al.}(2015)\citenamefont{Rudenko, Yuan, and Katsnelson}}]{PhysRevB.92.085419}
\bibinfo{author}{\bibfnamefont{A.~N.} \bibnamefont{Rudenko}}, \bibinfo{author}{\bibfnamefont{S.}~\bibnamefont{Yuan}}, \bibnamefont{and} \bibinfo{author}{\bibfnamefont{M.~I.} \bibnamefont{Katsnelson}}, \bibinfo{journal}{Phys. Rev. B} \textbf{\bibinfo{volume}{92}}, \bibinfo{pages}{085419} (\bibinfo{year}{2015}), \urlprefix\url{https://link.aps.org/doi/10.1103/PhysRevB.92.085419}.

\bibitem[{\citenamefont{Hu et~al.}(2024)\citenamefont{Hu, Xu, and Lian}}]{PhysRevB.110.L201106}
\bibinfo{author}{\bibfnamefont{Y.}~\bibnamefont{Hu}}, \bibinfo{author}{\bibfnamefont{Y.}~\bibnamefont{Xu}}, \bibnamefont{and} \bibinfo{author}{\bibfnamefont{B.}~\bibnamefont{Lian}}, \bibinfo{journal}{Phys. Rev. B} \textbf{\bibinfo{volume}{110}}, \bibinfo{pages}{L201106} (\bibinfo{year}{2024}), \urlprefix\url{https://link.aps.org/doi/10.1103/PhysRevB.110.L201106}.

\bibitem[{\citenamefont{Wen}(1990)}]{PhysRevB.42.6623}
\bibinfo{author}{\bibfnamefont{X.~G.} \bibnamefont{Wen}}, \bibinfo{journal}{Phys. Rev. B} \textbf{\bibinfo{volume}{42}}, \bibinfo{pages}{6623} (\bibinfo{year}{1990}), \urlprefix\url{https://link.aps.org/doi/10.1103/PhysRevB.42.6623}.

\bibitem[{\citenamefont{Shavit and Oreg}(2024)}]{shavit_quantum_2024}
\bibinfo{author}{\bibfnamefont{G.}~\bibnamefont{Shavit}} \bibnamefont{and} \bibinfo{author}{\bibfnamefont{Y.}~\bibnamefont{Oreg}}, \bibinfo{journal}{Physical Review Letters} \textbf{\bibinfo{volume}{133}}, \bibinfo{pages}{156504} (\bibinfo{year}{2024}), ISSN \bibinfo{issn}{0031-9007, 1079-7114}, \urlprefix\url{https://link.aps.org/doi/10.1103/PhysRevLett.133.156504}.

\bibitem[{\citenamefont{Shi et~al.}(2015)\citenamefont{Shi, Hong, Bechtel, Zeng, Martin, Watanabe, Taniguchi, Shen, and Wang}}]{Shi2015}
\bibinfo{author}{\bibfnamefont{Z.}~\bibnamefont{Shi}}, \bibinfo{author}{\bibfnamefont{X.}~\bibnamefont{Hong}}, \bibinfo{author}{\bibfnamefont{H.~A.} \bibnamefont{Bechtel}}, \bibinfo{author}{\bibfnamefont{B.}~\bibnamefont{Zeng}}, \bibinfo{author}{\bibfnamefont{M.~C.} \bibnamefont{Martin}}, \bibinfo{author}{\bibfnamefont{K.}~\bibnamefont{Watanabe}}, \bibinfo{author}{\bibfnamefont{T.}~\bibnamefont{Taniguchi}}, \bibinfo{author}{\bibfnamefont{Y.-R.} \bibnamefont{Shen}}, \bibnamefont{and} \bibinfo{author}{\bibfnamefont{F.}~\bibnamefont{Wang}}, \bibinfo{journal}{Nature Photonics} \textbf{\bibinfo{volume}{9}}, \bibinfo{pages}{515–519} (\bibinfo{year}{2015}), ISSN \bibinfo{issn}{1749-4893}, \urlprefix\url{http://dx.doi.org/10.1038/nphoton.2015.123}.

\bibitem[{\citenamefont{Wang et~al.}(2020)\citenamefont{Wang, Zhao, Shi, Wu, Zhao, Jiang, Watanabe, Taniguchi, Zettl, Zhou et~al.}}]{Wang2020}
\bibinfo{author}{\bibfnamefont{S.}~\bibnamefont{Wang}}, \bibinfo{author}{\bibfnamefont{S.}~\bibnamefont{Zhao}}, \bibinfo{author}{\bibfnamefont{Z.}~\bibnamefont{Shi}}, \bibinfo{author}{\bibfnamefont{F.}~\bibnamefont{Wu}}, \bibinfo{author}{\bibfnamefont{Z.}~\bibnamefont{Zhao}}, \bibinfo{author}{\bibfnamefont{L.}~\bibnamefont{Jiang}}, \bibinfo{author}{\bibfnamefont{K.}~\bibnamefont{Watanabe}}, \bibinfo{author}{\bibfnamefont{T.}~\bibnamefont{Taniguchi}}, \bibinfo{author}{\bibfnamefont{A.}~\bibnamefont{Zettl}}, \bibinfo{author}{\bibfnamefont{C.}~\bibnamefont{Zhou}}, \bibnamefont{et~al.}, \bibinfo{journal}{Nature Materials} \textbf{\bibinfo{volume}{19}}, \bibinfo{pages}{986–991} (\bibinfo{year}{2020}), ISSN \bibinfo{issn}{1476-4660}, \urlprefix\url{http://dx.doi.org/10.1038/s41563-020-0652-5}.

\bibitem[{\citenamefont{Harper}(1955)}]{harper_single_1955}
\bibinfo{author}{\bibfnamefont{P.~G.} \bibnamefont{Harper}}, \bibinfo{journal}{Proceedings of the Physical Society. Section A} \textbf{\bibinfo{volume}{68}}, \bibinfo{pages}{874} (\bibinfo{year}{1955}), ISSN \bibinfo{issn}{0370-1298}, \urlprefix\url{https://iopscience.iop.org/article/10.1088/0370-1298/68/10/304}.

\bibitem[{\citenamefont{Sokoloff}(1981)}]{sokoloff_band_1981}
\bibinfo{author}{\bibfnamefont{J.~B.} \bibnamefont{Sokoloff}}, \bibinfo{journal}{Physical Review B} \textbf{\bibinfo{volume}{23}}, \bibinfo{pages}{6422} (\bibinfo{year}{1981}), ISSN \bibinfo{issn}{0163-1829}, \urlprefix\url{https://link.aps.org/doi/10.1103/PhysRevB.23.6422}.

\bibitem[{\citenamefont{Hofstadter}(1976)}]{hofstadter_energy_1976}
\bibinfo{author}{\bibfnamefont{D.~R.} \bibnamefont{Hofstadter}}, \bibinfo{journal}{Physical Review B} \textbf{\bibinfo{volume}{14}}, \bibinfo{pages}{2239} (\bibinfo{year}{1976}), ISSN \bibinfo{issn}{0556-2805}, \urlprefix\url{https://link.aps.org/doi/10.1103/PhysRevB.14.2239}.

\bibitem[{\citenamefont{Sinner et~al.}(2023{\natexlab{b}})\citenamefont{Sinner, Pantaleón, and Guinea}}]{sinner_strain-induced_2023}
\bibinfo{author}{\bibfnamefont{A.}~\bibnamefont{Sinner}}, \bibinfo{author}{\bibfnamefont{P.~A.} \bibnamefont{Pantaleón}}, \bibnamefont{and} \bibinfo{author}{\bibfnamefont{F.}~\bibnamefont{Guinea}}, \bibinfo{journal}{Physical Review Letters} \textbf{\bibinfo{volume}{131}}, \bibinfo{pages}{166402} (\bibinfo{year}{2023}{\natexlab{b}}), ISSN \bibinfo{issn}{0031-9007, 1079-7114}, \urlprefix\url{https://link.aps.org/doi/10.1103/PhysRevLett.131.166402}.

\bibitem[{\citenamefont{Kohn and Sham}(1965)}]{Kohn1965}
\bibinfo{author}{\bibfnamefont{W.}~\bibnamefont{Kohn}} \bibnamefont{and} \bibinfo{author}{\bibfnamefont{L.~J.} \bibnamefont{Sham}}, \bibinfo{journal}{Phys. Rev.} \textbf{\bibinfo{volume}{140}}, \bibinfo{pages}{A1133} (\bibinfo{year}{1965}).

\bibitem[{\citenamefont{Kresse and Furthm\"{u}ller}(1996)}]{Kresse1996}
\bibinfo{author}{\bibfnamefont{G.}~\bibnamefont{Kresse}} \bibnamefont{and} \bibinfo{author}{\bibfnamefont{J.}~\bibnamefont{Furthm\"{u}ller}}, \bibinfo{journal}{Computational Materials Science} \textbf{\bibinfo{volume}{6}}, \bibinfo{pages}{15–50} (\bibinfo{year}{1996}), ISSN \bibinfo{issn}{0927-0256}, \urlprefix\url{http://dx.doi.org/10.1016/0927-0256(96)00008-0}.

\bibitem[{\citenamefont{Bl\"{o}chl}(1994)}]{Blochl1994}
\bibinfo{author}{\bibfnamefont{P.~E.} \bibnamefont{Bl\"{o}chl}}, \bibinfo{journal}{Phys. Rev. B} \textbf{\bibinfo{volume}{50}}, \bibinfo{pages}{17953} (\bibinfo{year}{1994}).

\bibitem[{\citenamefont{Kresse and Joubert}(1999)}]{Kresse1999}
\bibinfo{author}{\bibfnamefont{G.}~\bibnamefont{Kresse}} \bibnamefont{and} \bibinfo{author}{\bibfnamefont{D.}~\bibnamefont{Joubert}}, \bibinfo{journal}{Phys. Rev. B} \textbf{\bibinfo{volume}{59}}, \bibinfo{pages}{1758} (\bibinfo{year}{1999}).

\bibitem[{\citenamefont{Perdew et~al.}(1996)\citenamefont{Perdew, Burke, and Ernzerhof}}]{Perdew1996}
\bibinfo{author}{\bibfnamefont{J.~P.} \bibnamefont{Perdew}}, \bibinfo{author}{\bibfnamefont{K.}~\bibnamefont{Burke}}, \bibnamefont{and} \bibinfo{author}{\bibfnamefont{M.}~\bibnamefont{Ernzerhof}}, \bibinfo{journal}{Phys. Rev. Lett.} \textbf{\bibinfo{volume}{77}}, \bibinfo{pages}{3865} (\bibinfo{year}{1996}).

\bibitem[{\citenamefont{Pan et~al.}(2017)\citenamefont{Pan, Wang, Xiao, Hu, and Yao}}]{PhysRevB.96.041411}
\bibinfo{author}{\bibfnamefont{D.}~\bibnamefont{Pan}}, \bibinfo{author}{\bibfnamefont{T.-C.} \bibnamefont{Wang}}, \bibinfo{author}{\bibfnamefont{W.}~\bibnamefont{Xiao}}, \bibinfo{author}{\bibfnamefont{D.}~\bibnamefont{Hu}}, \bibnamefont{and} \bibinfo{author}{\bibfnamefont{Y.}~\bibnamefont{Yao}}, \bibinfo{journal}{Phys. Rev. B} \textbf{\bibinfo{volume}{96}}, \bibinfo{pages}{041411} (\bibinfo{year}{2017}), \urlprefix\url{https://link.aps.org/doi/10.1103/PhysRevB.96.041411}.

\bibitem[{\citenamefont{Grimme et~al.}(2011)\citenamefont{Grimme, Ehrlich, and Goerigk}}]{grimme2011effect}
\bibinfo{author}{\bibfnamefont{S.}~\bibnamefont{Grimme}}, \bibinfo{author}{\bibfnamefont{S.}~\bibnamefont{Ehrlich}}, \bibnamefont{and} \bibinfo{author}{\bibfnamefont{L.}~\bibnamefont{Goerigk}}, \bibinfo{journal}{J. Comput. Chem.} \textbf{\bibinfo{volume}{32}}, \bibinfo{pages}{1456} (\bibinfo{year}{2011}).

\bibitem[{\citenamefont{Hamada}(2014)}]{revvdwdf2}
\bibinfo{author}{\bibfnamefont{I.}~\bibnamefont{Hamada}}, \bibinfo{journal}{Phys. Rev. B} \textbf{\bibinfo{volume}{89}}, \bibinfo{pages}{121103} (\bibinfo{year}{2014}).

\bibitem[{\citenamefont{Smidstrup et~al.}(2019)\citenamefont{Smidstrup, Markussen, Vancraeyveld, Wellendorff, Schneider, Gunst, Verstichel, Stradi, Khomyakov, Vej-Hansen et~al.}}]{Smidstrup2019}
\bibinfo{author}{\bibfnamefont{S.}~\bibnamefont{Smidstrup}}, \bibinfo{author}{\bibfnamefont{T.}~\bibnamefont{Markussen}}, \bibinfo{author}{\bibfnamefont{P.}~\bibnamefont{Vancraeyveld}}, \bibinfo{author}{\bibfnamefont{J.}~\bibnamefont{Wellendorff}}, \bibinfo{author}{\bibfnamefont{J.}~\bibnamefont{Schneider}}, \bibinfo{author}{\bibfnamefont{T.}~\bibnamefont{Gunst}}, \bibinfo{author}{\bibfnamefont{B.}~\bibnamefont{Verstichel}}, \bibinfo{author}{\bibfnamefont{D.}~\bibnamefont{Stradi}}, \bibinfo{author}{\bibfnamefont{P.~A.} \bibnamefont{Khomyakov}}, \bibinfo{author}{\bibfnamefont{U.~G.} \bibnamefont{Vej-Hansen}}, \bibnamefont{et~al.}, \bibinfo{journal}{Journal of Physics: Condensed Matter} \textbf{\bibinfo{volume}{32}}, \bibinfo{pages}{015901} (\bibinfo{year}{2019}), ISSN \bibinfo{issn}{1361-648X}, \urlprefix\url{http://dx.doi.org/10.1088/1361-648X/ab4007}.

\bibitem[{\citenamefont{Kane et~al.}(2017)\citenamefont{Kane, Stern, and Halperin}}]{PhysRevX.7.031009}
\bibinfo{author}{\bibfnamefont{C.~L.} \bibnamefont{Kane}}, \bibinfo{author}{\bibfnamefont{A.}~\bibnamefont{Stern}}, \bibnamefont{and} \bibinfo{author}{\bibfnamefont{B.~I.} \bibnamefont{Halperin}}, \bibinfo{journal}{Phys. Rev. X} \textbf{\bibinfo{volume}{7}}, \bibinfo{pages}{031009} (\bibinfo{year}{2017}), \urlprefix\url{https://link.aps.org/doi/10.1103/PhysRevX.7.031009}.

\bibitem[{\citenamefont{Teo and Kane}(2014)}]{PhysRevB.89.085101}
\bibinfo{author}{\bibfnamefont{J.~C.~Y.} \bibnamefont{Teo}} \bibnamefont{and} \bibinfo{author}{\bibfnamefont{C.~L.} \bibnamefont{Kane}}, \bibinfo{journal}{Phys. Rev. B} \textbf{\bibinfo{volume}{89}}, \bibinfo{pages}{085101} (\bibinfo{year}{2014}), \urlprefix\url{https://link.aps.org/doi/10.1103/PhysRevB.89.085101}.

\end{thebibliography}
%+++++++++++++++++++++++++++++++++++++++++++++++++++++++++++++++++++++

%\begin{thebibliography}{43}
%\expandafter\ifx\csname natexlab\endcsname\relax\def\natexlab#1{#1}\fi
%\expandafter\ifx\csname bibnamefont\endcsname\relax
%  \def\bibnamefont#1{#1}\fi
%\expandafter\ifx\csname bibfnamefont\endcsname\relax
%  \def\bibfnamefont#1{#1}\fi
%\expandafter\ifx\csname citenamefont\endcsname\relax
%  \def\citenamefont#1{#1}\fi
%\expandafter\ifx\csname url\endcsname\relax
%  \def\url#1{\texttt{#1}}\fi
%\expandafter\ifx\csname urlprefix\endcsname\relax\def\urlprefix{URL }\fi
%\providecommand{\bibinfo}[2]{#2}
%\providecommand{\eprint}[2][]{\url{#2}}
%
%
%\bibitem[{\citenamefont{Tombros et~al.}(2007)\citenamefont{Tombros, Jozsa,
%  Popinciuc, Jonkman, and van Wees}}]{gra-spin1}
%\bibinfo{author}{\bibfnamefont{N.}~\bibnamefont{Tombros}},
%  \bibinfo{author}{\bibfnamefont{C.}~\bibnamefont{Jozsa}},
%  \bibinfo{author}{\bibfnamefont{M.}~\bibnamefont{Popinciuc}},
%  \bibinfo{author}{\bibfnamefont{H.~T.} \bibnamefont{Jonkman}},
%  \bibnamefont{and} \bibinfo{author}{\bibfnamefont{B.~J.} \bibnamefont{van
%  Wees}}, \bibinfo{journal}{Nature} \textbf{\bibinfo{volume}{448}},
%  \bibinfo{pages}{571} (\bibinfo{year}{2007}).
%
%
%\end{thebibliography}

%\bibitem{Snote}
%\bibinfo{note}{See Supplemental Material at %http://link.aps.org/supplemental/???
%for details for DFT and transport calculations, and crystal and electronic structures, Rashba angles, and charge to spin conversion results of all twist angles, Supplemental Notes S1-S2, Supplemental Table S1, and Supplemental Figs. S1–S7.}

\end{document}